\documentclass{article}

\usepackage[final,main]{neurips_2026}

\usepackage[utf8]{inputenc} 
\usepackage[T1]{fontenc}    
\usepackage{booktabs}       
\usepackage{amsfonts}       
\usepackage{amsmath}
\usepackage{nicefrac}       
\usepackage{microtype}      
\usepackage{xcolor}         
\usepackage{graphicx}

\usepackage{array}
\usepackage{multirow}
\usepackage{wrapfig}
\usepackage{subcaption}

\usepackage{hyperref}       
\usepackage{url}            

\definecolor{softyellow}{RGB}{255, 249, 196}
\definecolor{softorange}{RGB}{255, 224, 178}

\definecolor{softblue}{RGB}{187, 222, 251}
\definecolor{softcyan}{RGB}{178, 235, 242} 

\definecolor{softgreen}{RGB}{200, 230, 201}
\definecolor{softmint}{RGB}{178, 223, 219}

\definecolor{darkblue}{rgb}{0, 0, 0.5}
\hypersetup{colorlinks=true, citecolor=darkblue, linkcolor=darkblue, urlcolor=darkblue}

\title{Voice \textit{``Cloning''} is Style Transfer}

\author{%
  Kaitlyn Zhou\textsuperscript{\textnormal{1,2}}\qquad
  Federico Bianchi\textsuperscript{\textnormal{2}}\qquad
  Martijn Bartelds\textsuperscript{\textnormal{2}} \\
  \textbf{Anna Pot}\textsuperscript{3}\qquad
  \textbf{Yongchan Kwon}\textsuperscript{2}\qquad
  \textbf{James Zou}\textsuperscript{2,3} \\
  \textsuperscript{1}Cornell University \\
  \textsuperscript{2}TogetherAI \\
  \textsuperscript{3}Stanford University \\
  \texttt{kaitlynz@cornell.edu}
}

\begin{document}

\maketitle
\begin{abstract}
Artificially generated speech is increasingly embedded in everyday life. Voice cloning in particular enables applications where identity preservation is important, such as completing a recording, dubbing in a new language, or preserving the voices of individuals with speech loss. However, in our work, we find that despite the term, voice cloning does not faithfully \textit{``clone''} an individual's voice. Instead, we find that widely-used voice cloning models systematically apply style transfer to source voices. As rated by human annotators, cloned voices are perceived as more authoritative, warm, customer-service-like, and human-like compared to their sources. Human annotators also report greater trust in cloned voices than source voices, and a greater willingness to disclose sensitive personal information to them. Our work furthermore shows that voice cloning leads to homogenization of speaker characteristics, as measured by reduced variance in accent, speaking rate, and the audio embedding space. Together, our results highlight a new set of limitations and risks of voice cloning technology and their potential impact on human behavior.
\end{abstract}
\section{Introduction}
A person’s voice is a deeply personal marker of identity, conveying accent, affect, and individual character. Recent advances in text-to-speech (TTS) have enabled increasingly humanlike speech generation, with zero-shot voice cloning emerging as a prominent application. Most discourse around voice cloning has focused on harms in misuse: a more faithful clone can make impersonation, fraud, and other unauthorized uses more convincing. However, in settings where users intentionally provide their voices --- such as enhanced presentations \citep{zheng2025learning} and pronunciations \citep{park_esl_clone}, assistive technologies for individuals with speech impairments \citep{wairagkar2025instantaneous, chavan_chi_ea_aac}, generative tools for personal well-being and self-reflection \citep{pataranutaporn2021ai}, and multilingual media dubbing \citep{li-etal-2024-pause} --- vocal fidelity is precisely what gives the technology its value. In these cases, unfaithful cloning can also be harmful: it may distort how a person sounds, erase identity-linked vocal traits, or replace their voice with a more standardized version.

Although these systems are often described as \textit{``cloning''} a voice, it remains unclear whether they actually preserve the speaker identity they claim to reproduce. We examine this assumption empirically and find that modern voice cloning systems fall short of faithful reproduction. Rather than preserving source voices, they systematically transform them, diminishing distinctive vocal traits and replacing them with more homogenized, socially preferred speaking characteristics.

Our study presents two key findings regarding modern voice cloning systems. First, we show how voice \textit{``cloning''} systematically applies style transformations to source voices. These style transfers arise without explicit authorization from users and reflect limitations in voice cloning fidelity. Through annotations of paired source and cloned recordings, we observe that cloned voices sound significantly warmer, more authoritative, and more customer-service-oriented. Cloned voices in our study are often perceived as more \textit{``human-like,''} contributing to a phenomenon akin to hyperrealism --- previously documented in visual generative models but not yet widely observed in speech \citep{hyperrealism-ai-miller, lavan2025voice}. These vocal transformations influence more than just human perceptions: our behavioral evaluations indicate that listeners report greater trust in cloned voices and express higher willingness to engage in intimate conversations with them, with direct consequences for downstream human behaviors. This finding underscores the need for continued efforts to assess the downstream effects of synthetic voice.

Second, voice cloning homogenizes identity. It systematically imposes a \textit{``native''} English accent, collapsing the variation present in non-native English speech and erasing markers of cultural origin and individual distinctiveness. Using speaker identity probe methods, we find that cloned voices lose the features that make source voices distinct from one another, with more identity confusion in cloned outputs relative to source recordings. This stylistic transformation compounds over time. Repeated cloning produces directional drift in audio embedding space, increased pitch in recordings, and altered emotional expression -- all of which further distance outputs from the source speaker.

Our findings contribute to a deeper technical and interactional understanding of voice cloning, with direct implications for how these systems are deployed and governed. Our results highlight the need to transparently reveal systematic transformations introduced by these models---such as stylistic shifts and homogenization---that shape how voices are perceived and interpreted. This adds to a long line of work that discusses risks and limitations of voice cloning, such as impersonation, fraud, bypassing voice-based authentication \citep{not_by_voice_hutiri, cscw_du_voice_cloning}, labor displacement of workers whose voices and livelihoods these systems increasingly replace \citep{almeda2025labor}, unauthorized use of voices \citep{agnew2024sound}, and loss of personal identity \citep{leuenberger2025role}. Taken together, this work suggests that deviations from faithful voice reproduction are not merely technical shortcomings but a new form of homogenization with implications for consumer protection, cultural diversity, and personal identity.

The wide-scale deployment of voice cloning technology is not inevitable but reflects choices made by companies, policymakers, and researchers \citep{bijker1987social, winner2017artifacts}. Our work aims to provide a clear empirical understanding of what these systems actually do to inform grounded technology design and policy responses.

\section{Related Work}
Research on voice cloning is part of a broader literature that includes speech synthesizers and text-to-speech systems \citep{arik2018neuralvoicecloningsamples,le2023voiceboxtextguidedmultilingualuniversal,wang2023neuralcodeclanguagemodels}. TTS is commonly used in speech technologies for synthetic voice generation, and voice cloning extends the TTS architecture. Voice cloning extracts a speaker-specific embedding from reference audio and incorporates it into the generation pipeline, thereby enabling the system to mimic a particular individual’s voice. 

Prior work primarily emphasized a model's ability to produce ``natural'' and humanlike voice \citep{shen2023naturalspeech2latentdiffusion, ju2024naturalspeech3zeroshotspeech, anastassiou2024seedttsfamilyhighqualityversatile}. Meanwhile, other works have tried to add more control to the way zero-shot voice cloning is done by controlling aspects like the timbre, content, and style of a voice \citep{ji-etal-2025-controlspeech}, the emotions a voice conveys \citep{chen2024emoknobenhancevoicecloning}, and even controlling the accent, rhythm, and intonation \citep{qin2024openvoiceversatileinstantvoice}. At times, this involves using natural language prompts to direct the style of the generation \citep{promptts, yang2023instructttsmodellingexpressivetts}. 

Despite rapid technical progress, voice cloning remains contested due to concerns about misuse, which include impersonation, fraud, and bypassing voice-based authentication \citep{not_by_voice_hutiri, cscw_du_voice_cloning}, the unauthorized use of voices in datasets \citep{agnew2024sound}, and exaggeration of accents \citep{accent_bias_facct}. Studies from as early as 2023 show that audio deepfakes are indistinguishable from real voices \citep{mai2023warning}. Work in Human-Computer Interaction (HCI) has shown the economic impact on voice actors \citep{almeda2025labor}. Philosophical work has argued that cloned voices function as simulations detached from their embodied origins \citep{berkowitz2026simulating} and risk the loss of personal and cultural identity \citep{leuenberger2025role}. Legal challenges in voice cloning are complex \citep{lee2026vocal, berkowitz2025look}, and some scholars have noted the implications of false light\footnote{\url{https://www.law.cornell.edu/wex/false_light}} and defamation that could be applied to cloned voices \citep{wells2022s}. To this end, advances in natural language processing have made efforts to preserve the privacy of users while still leveraging the advances of voice cloning. For example, \citet{platnick-etal-2024-preset} discusses using voice cloning technology by closely matching the input of a voice to a similar preset consenting voice for cloning. Other design practices, such as those in South Korea, aim to explicitly mark generated AI voices for transparency to older adult customers \cite{nytimes2026koreaAI}. 

\section{Methods}

\begin{figure}[h!]
    \centering
    \includegraphics[width=0.88\linewidth]{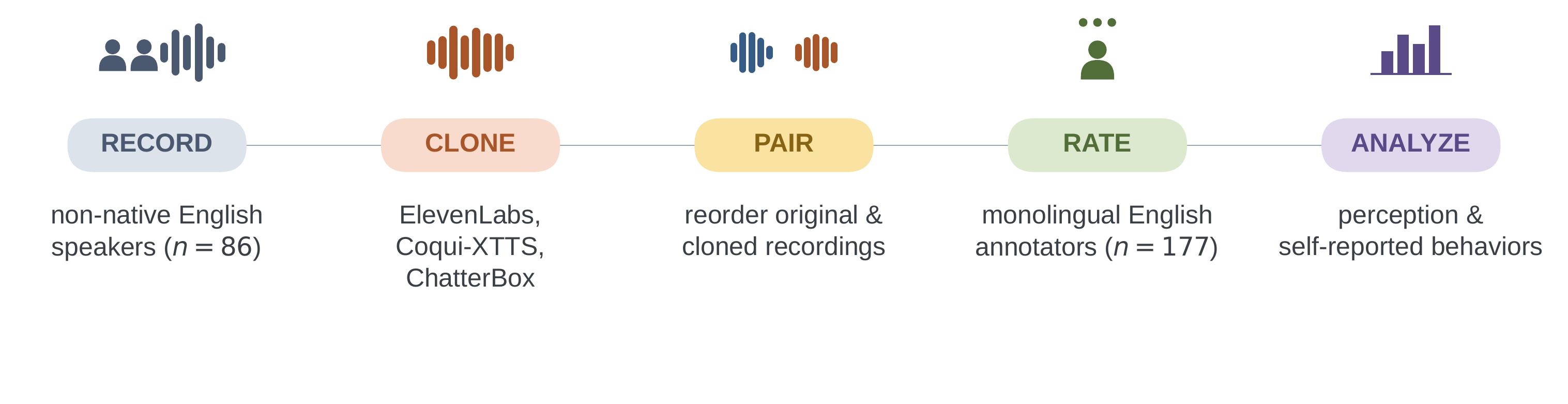}
    \caption{\textbf{Study pipeline.} Audio from $n{=}86$ non-native English speakers serves as reference for cloning on three models (\textsc{ElevenLabs}, \textsc{Coqui-XTTS}, \textsc{ChatterBox}). Each source recording is paired with its clone and shown in randomized order to $n{=}177$ annotators.}
    \label{fig:placeholder}
\end{figure}

\subsection{Audio Data Collection}

\begin{wraptable}{r}{0.4\textwidth}
\vspace{-3em}
\centering
\small
\caption{Speaker demographics ($N=86$).}
\label{tab:speaker-demographics}
\begin{tabular}{@{}lc@{}}
\toprule
\textbf{Variable} & \textbf{Value} \\
\midrule
\multicolumn{2}{@{}l}{\textit{Sex}} \\
\quad Male & 43 (50.0\%) \\
\quad Female & 43 (50.0\%) \\
\midrule
\multicolumn{2}{@{}l}{\textit{Age (years)}} \\
\quad Mean\,$\pm$\,SD & 38.3\,$\pm$\,10.7 \\
\quad Range & 19--64 \\
\midrule
\multicolumn{2}{@{}l}{\textit{Self-reported accent}} \\
\quad Mean\,$\pm$\,SD & 3.8\,$\pm$\,3.2 \\
\quad Range & 0--10 \\
\bottomrule
\end{tabular}
\end{wraptable}
The experimental setup begins by recruiting $n=86$ participants (over 18 years old, based in the U.S.) via Prolific. Our research goal is to understand how these voice cloning systems might impact a broad user population, thus we wanted our speakers to reflect this diversity. We recruit non-Native English speakers who have a wide range of foreign accents, some self-reporting none, while others report strong accents. Participants also vary in age and are sex balanced, Table \ref{tab:speaker-demographics}. Participants started by providing information about their English language background, and then were asked to read aloud the Grandfather Passage \citep{van1972speech}, a nine-sentence standard text widely used in speech and language assessment. 

Each recording was automatically split into sentence-level audio clips using Whisper-based forced alignment \citep{radford2023robust}, yielding up to nine clips per participant. We conducted manual quality control on each sentence clip, reviewing alignment boundaries and ensuring every clip was read correctly. Clips containing incorrect utterances were excluded. Because speakers varied in reading accuracy and speaking rate, and because recordings were capped at 90 seconds, some participants contributed fewer valid clips than others. To preserve speaker diversity, we retained all valid utterances from speakers rather than requiring a complete set of nine sentences. The final dataset contains $86$ speakers and $699$ valid sentence-level clips across the nine target sentences.

Audio clips were preprocessed by trimming silent segments and normalizing amplitude to a consistent level, common in speech processing pipelines \citep{labied2022overview, keerio2009preprocessing}. Upon publication, we will make this dataset available via Hugging Face for research purposes only. This study was IRB-approved, and all participants were paid \$18 USD per hour. Details of the consent form, questionnaire, and full passage can be found in \S\ref{section:methods_details}, along with how we protect participant privacy in \S\ref{sec:privacy_terms} and task screenshots (Figure \ref{fig:annotation_screenshot}).

\subsection{Voice Cloning}
We evaluate three widely used TTS models --- two open-source (\texttt{ChatterBox}, \texttt{Coqui-XTTS}) and one state-of-the-art proprietary model \citep[\texttt{ElevenLabs V3},][]{chatterboxtts2025, Eren_Coqui_TTS_2021, elevenlabs_v3_2026}. Open-source models were selected to reduce privacy risks by enabling greater control over speaker data, while \texttt{ElevenLabs} was included as a leading proprietary system that provides mechanisms for data removal and opt-out from model training. The dataset and code are available at the following links.\footnote{\url{https://huggingface.co/datasets/kzhou/voice_cloning_style_transfer}}\footnote{\url{https://github.com/kzhou-cloud/voice-cloning-public}}

Our unit of analysis is a \emph{tuple} $(s, \ell)$, where $s$ denotes the speaker and $\ell \in \{1,\ldots,9\}$ indexes the sentence of the passage. Each tuple contains two audio samples: a human-produced source and a model-generated clone of the same utterance.

\begin{figure}[]
\centering
\includegraphics[width=0.75\linewidth]{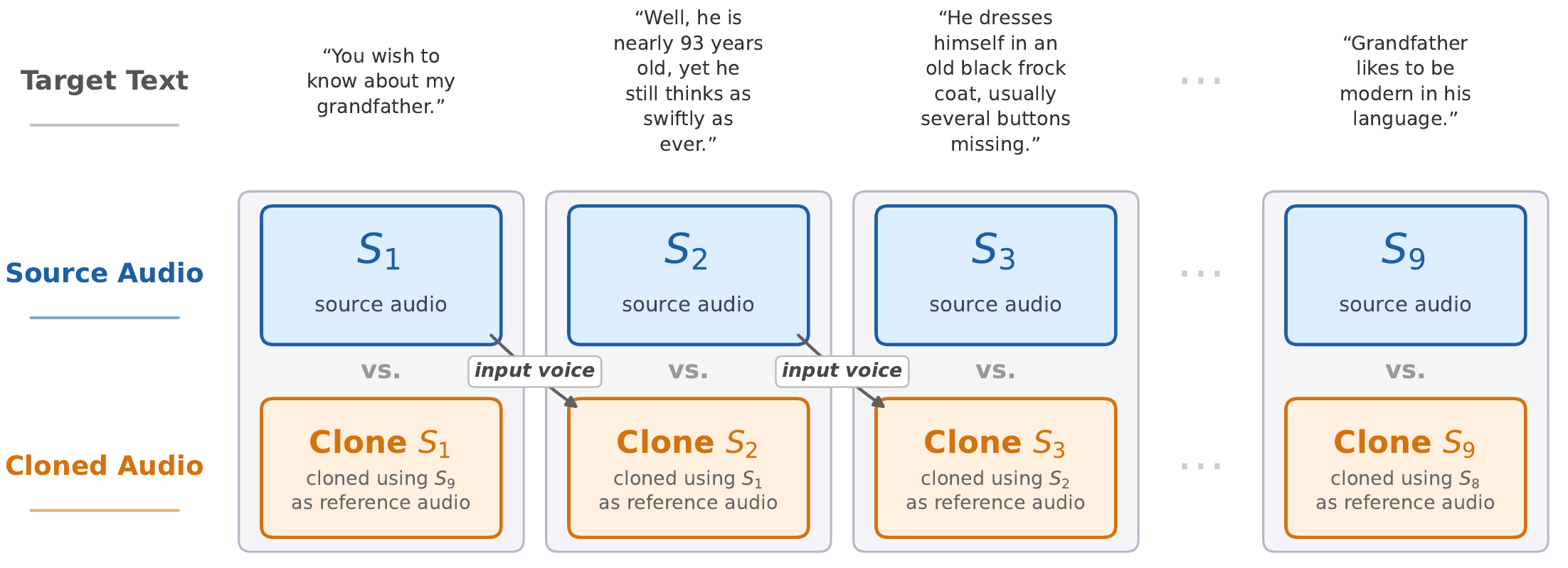}
\caption{Illustration of \textit{cross-sentence} voice cloning.}
\label{fig:placeholder}
\end{figure}

We generate voice clones using a \textit{cross-sentence} cloning paradigm. In each cross-sentence pair, the model takes in $S_\ell ^{\mathrm{source}}$ as the reference audio and generates $S_{\ell+1}^{\mathrm{cloned}}$, using the target text of sentence ${\ell+1}$. The pairs wrap cyclically ($S_9^{\mathrm{source}} \rightarrow S_1^{\mathrm{cloned}}$). This offset of voice cloning ensures that the speech model isn't given speaker characteristics for the target cloned audio, but instead needs to extract generalizable speaker features from the reference audio. In human annotation evaluation, we realign source audio with cloned audio so that we are always comparing $S_{\ell+1}^{\mathrm{cloned}}$ with $S_{\ell+1}^{\mathrm{source}}$. 

\subsection{Voice Annotation}
\label{sec:human-eval}
We present the source and cloned recordings to online annotators to examine how human listeners characterize both source and cloned speech recordings. Each annotation session contains clips from 10 distinct $(s,\ell)$ tuples, 10 source recordings and their 10 cloned counterparts. The 20 clips are then \emph{globally shuffled}, so annotators are unaware of whether any given clip is human-produced or generated. For every clip, annotators provided ratings on five-point Likert-scale dimensions, including human-likeness, customer-service voice quality, authoritativeness, warmth, trustworthiness, and suitability for intimate conversation, see rationales in Table \ref{tab:dimensions}.

To control for potential confounding factors, each session is limited to one TTS model and one speaker-sex category. The rationale is that one model may sound substantially more authoritative than another, so combining clips from different models could introduce interaction effects in annotator judgments. In that case, it would be difficult to determine whether observed differences reflect the distinction between source and cloned audio or differences between the models themselves. The same concern applies to speaker sex: female voices could be rated differently from male voices, mixing female and male voices within a session could therefore similarly influence annotation outcomes.

Annotators were recruited via Prolific, with eligibility restricted to U.S.-based participants who reported English as their only language, to minimize potential ingroup–outgroup confounds between listeners and annotators. Each annotator could complete up to ten independent sessions. In total, we collected 4,000 annotations from 177 unique participants across four conditions: the three systems above, plus an \texttt{ElevenLabs} low-expressiveness ablation (\S\ref{sec:findings1}). The analyses that compare the three systems therefore use the 3{,}000 annotations from 137 of those participants. 

\begin{table}[h!]
\centering\small
\begin{tabular}{
  >{\raggedright\arraybackslash}p{0.18\linewidth}%
  >{\raggedright\arraybackslash}p{0.34\linewidth}%
  >{\raggedright\arraybackslash}p{0.34\linewidth}%
}
\toprule
\textbf{Dimension} & \textbf{Main Question} & \textbf{Rationale} \\
\midrule
\texttt{humanlike} &
\textit{Does the clip sound humanlike?} &
Measuring if cloned voices were as humanlike as source voices (intentionally ambiguous)\\
\texttt{customer service} &
\textit{Does the clip have a ``customer service voice''?} &
Measuring professional/assistant-like register (e.g., polished, deferential, scripted service tone). \\
\texttt{authoritative} &
\textit{Does the clip sound authoritative?} &
Measuring perceptions of warmth and competence, adjusted for speech  \citep{fiske2007universal}. \\
\texttt{warm} &
\textit{Does the clip sound warm?} &
Same as above\\
\midrule
\texttt{trust} &
\textit{Would you trust instructions or answers from this voice?} &
Measuring self-reported reliance and trust behaviors.\\
\texttt{intimate} &
\textit{Would you be comfortable having an intimate conversation with this voice?} &
Measuring self-reported willingness to disclose personal information.\\
\midrule
\texttt{native\_english} &
\textit{Does the clip above sound like a native English speaker?} &
Measuring perceived accent transfers. Stereotypical (exaggerated) accent generation known in TTS \citep{accent_bias_facct}. \\
\bottomrule
\end{tabular}
\caption{Likert items rated per clip (scale $1$--$5$). See figure \ref{fig:annotation_screenshot} for details.}
\label{tab:dimensions}
\end{table}

\section{Voice Cloning as Style Transfer}
\label{sec:findings1}
Our first major finding is that voice cloning does not faithfully preserve speaker identity. We show that widely used voice-cloning systems often behave less like identity-preserved copying, and more like style-normalized transformation. 

\begin{figure}[h!]
 \centering
 \includegraphics[width=0.75\linewidth]{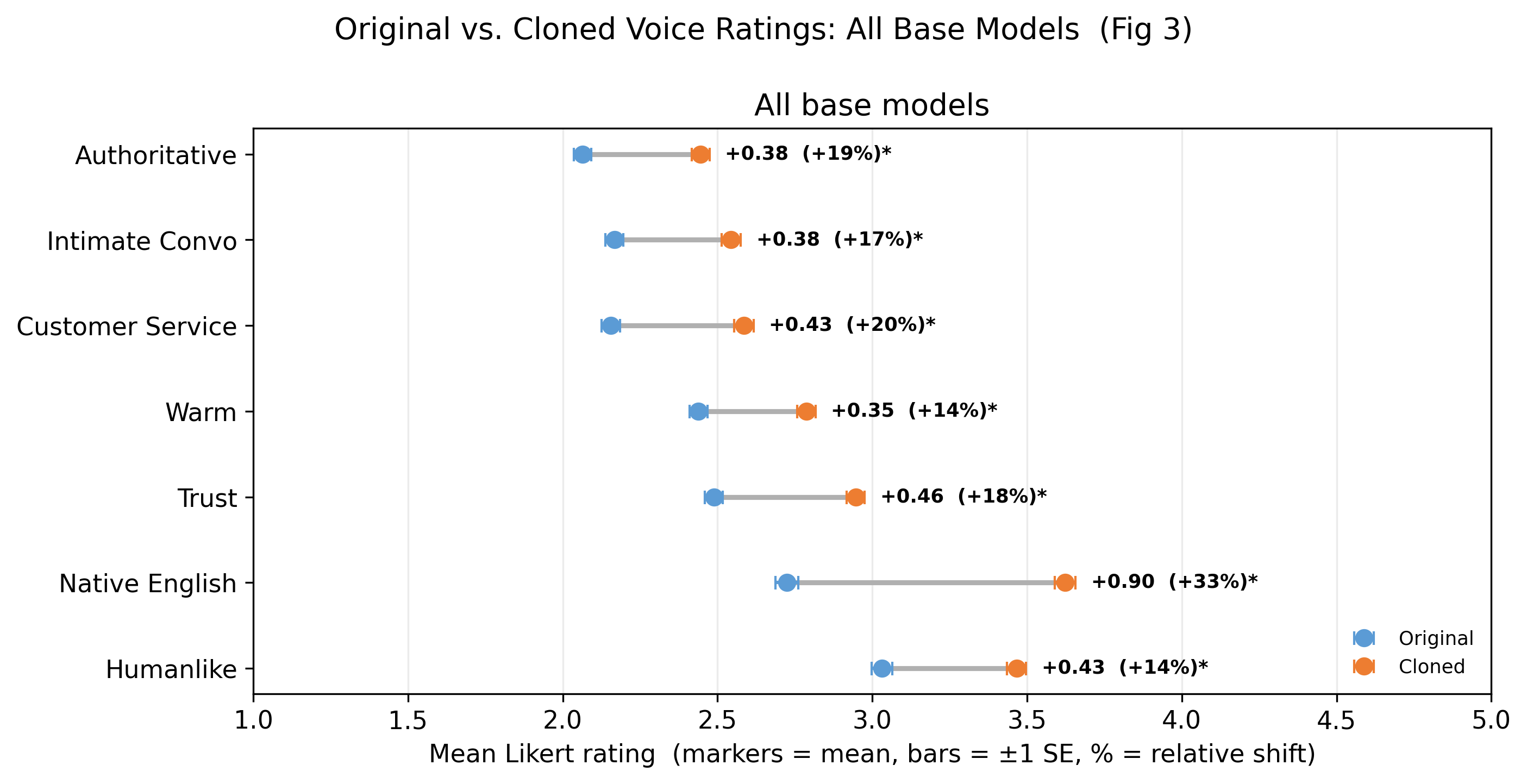}
 \caption{Rating differences between cloned and source voices across all three models (\texttt{ChatterBox}, \texttt{Coqui-XTTS}, \texttt{ElevenLabs V3}), standard error shown. All differences significant at 95\% by permutation test ($p<0.05$).}
 \label{fig:ratings}
\end{figure}

We find that cloned voices are seen as significantly more humanlike, more authoritative, more warm, and more customer-service-like. On a scale of 1 - 5, with an average score of 5 indicating that all annotators rated the audio files as "extremely warm", the source recordings had an average score of 2.4, while cloned recordings had a mean score of 2.8. In Figure \ref{fig:ratings}, we show that the difference between the two distributions is statistically significant at the 95\% confidence level based on the permutation test. We present the results in aggregate, but they are significant for each individual model as well. Because male and female voices differ substantially in speech acoustics \citep{titze1989physiologic, smith2007discrimination}, we further stratify our analysis by speaker sex,\footnote{We avoid collecting additional demographic information from participants and use speaker sex for demographic analysis, as provided by our annotation platform. We recognize the limitations and exclusionary nature of this classification and aim to have more inclusive measures of speaker identity for future work.} and again the effects persist. See additional figures in the appendix, Figure \ref{fig:ratings_by_model} and \ref{fig:ratings_by_speaker_sex}. We analyze the shift toward perceived \textit{``native''} English speech separately in Section~\ref{sec:findings2}. 

For additional robustness, we also fit a linear mixed-effects model: \texttt{rating $\sim$ cloned $+$ (1|speaker) $+$ (1|annotator) $+$ (1|sentence)}, with the TTS model as a fixed effect. This model accounts for our set-up where every speaker is rated by many annotators and every annotator rates many speakers. We also add per-clip audio-quality co-variates (z-scored duration, RMS energy, spectral centroid, and pitch).\footnote{\texttt{librosa.pyin} and \texttt{librosa.yin} were used to calculate pitch} Under both robustness checks our results hold as every trait remains significant with a 95\% confidence interval excluding zero ($n=3{,}000$ ratings, 86 speakers, 137 annotators; all $p<0.001$), Appendix Tables \ref{tab:mixed_effects} and \ref{tab:mixed_effects_bymodel}. 

\paragraph{Is the Perceptual Shift Just An Accent Effect?}
Since the perception of accent changed most dramatically, we wanted to understand if the other perception effects we saw were just a downstream result from the accent shift  (\S\ref{sec:findings2}). Here, we control for accent shift by restricting our dataset to speakers who self-report no foreign accent at all ($n=16$), and we find that the cloned voices of these speakers are still rated significantly more humanlike, warmer, more customer-service-like, more trustworthy, more authoritative, and more native-sounding (all $p<0.05$); only willingness to have an intimate conversation falls short of significance in this subgroup. If we increase to speakers who reported no minimal accent ($n=30$), every trait is significant. Additionally, we conducted a mediation analysis using perceived English nativeness as the mediator. The direct effect of cloning holds for every trait we tested except warmth, whose gain is largely accounted for by the nativeness shift, Appendix~\ref{sec:accent_controls}.

We also asked participants to indicate behavioral intentions upon listening to the source and cloned voices. Specifically, we asked how much they would trust answers and responses from the voices and how willing they would be to have intimate conversations with voices. In both cases, we see a significant increase in self-reported trust and in willingness to have an intimate conversation in response to cloned voices compared to source voices. Again, these are stated perceptions and intentions elicited in a rating task rather than real self-incentivized behavior. Future work necessitates in-situ evaluations with self-incentivized users performing tasks that require reliance and information disclosure \citet{zhou-etal-2025-rel}. 

Our findings are in line with work from \citep{nightingale} that shows that generated faces are seen as more trustworthy than real human faces, but a departure from prior work which found text-to-speech technology to be less trustworthy, \citep{chi_audio_less_trustworthy}. Our findings also complement the findings of \citep{mogi_chatbot_self_disclosure}, who found that engagement and disclosure are higher among one's own cloned voice rather than the cloned voices of family members or strangers. We discuss these findings and their implications in detail in \S\ref{sec:harms}. 

\subsection{Ablations}
To better understand why the cloning process may fail to preserve the reference speaker’s unique vocal characteristics, we examine two plausible mechanisms. One possible explanation is that \textbf{clip duration} affects fidelity, such that the source clips may be too short for the model to reliably capture speaker-specific characteristics. To test this, we run an ablation where we concatenate the first seven sentences (average of 37 seconds) and use these to generate the eighth. Figure \ref{fig:long_vs_short} shows the same style transformations being reproduced.

A second possibility is that the observed transformations arise from the default \textbf{generation settings}, which can be tuned. \texttt{ElevenLabs} and \texttt{Chatterbox} provide similarity-to-speaker and expressiveness controls.\footnote{Temperature is also a tunable parameter, but leads more easily to degenerate behavior at non-optimal settings; to be investigated in future work.} When comparing these settings to the defaults, we find that the defaults correspond to low expressiveness and high similarity in embedding space, suggesting that the generated voices are already highly similar to the source in terms of speaker fidelity, Figure \ref{fig:embedding_styles}. Evaluating \texttt{ElevenLabs} at low expressiveness also reproduces our results exactly, showing that these style transformations persist even at minimal expressiveness, see Figure \ref{fig:elevenlabs_low_expressiveness}.
\section{Voice Cloning as Homogenization}
 \label{sec:findings2}
In our results above, we illustrate that voice \textit{``cloning''} is not actually faithfully capturing a speaker's speech characteristics, but that a number of stylistic transformations are being applied. Here, we show that these stylistic transformations are not arbitrary, but are actually converging towards a particular way of speaking, a form of voice homogenization. 

\subsection{Homogenization in Accent and Cadence}

\begin{figure}[]
 \centering
 \includegraphics[width=0.82\linewidth]{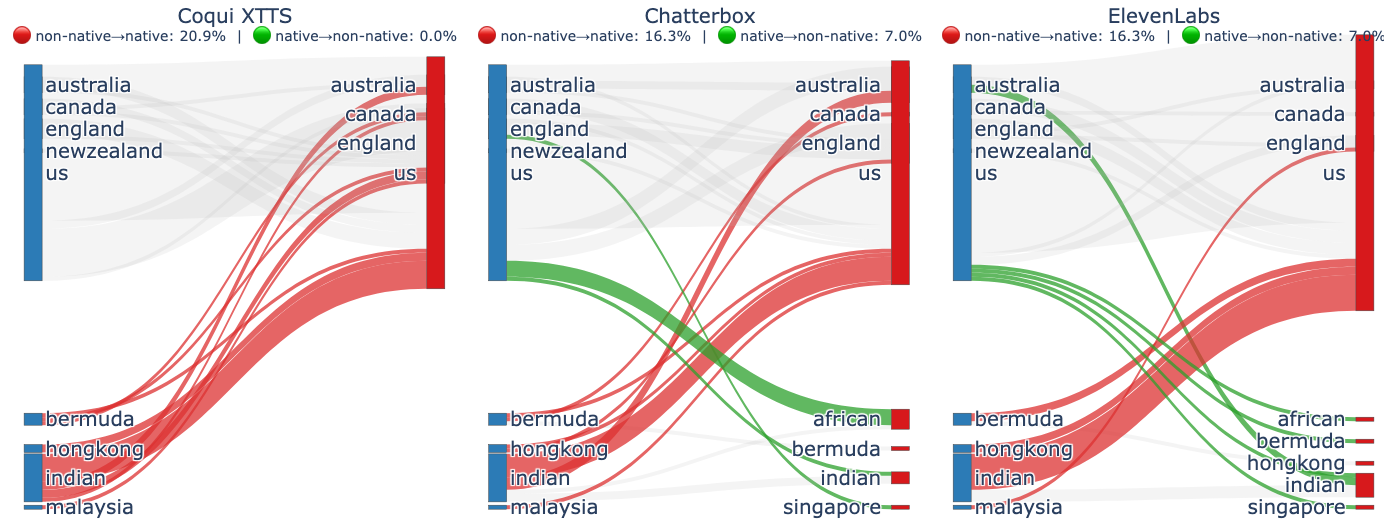}
 \caption{Shifts in classified accent after cloning. Sankey diagrams map source accent labels (left) to post-cloning labels (right). Red flows mark shifts toward dominant Anglophone varieties (U.S., U.K., Canadian, Australian, New Zealand), green the reverse, and gray labels that stay within the same broad category.}
 \label{fig:homogenization_accent}
\end{figure}

One of the biggest differences between source audio and cloned audio is that cloned audio has a dramatic increase in the perceived English \textit{``nativeness''} of the speech. Our non-native English speakers come from 22 unique language backgrounds, but their voice clones are judged to be significantly more similar to English \textit{``native''} speech than their original voices. To examine this shift at a more granular level, we use \texttt{CommonAccent} \citep{zuluaga2023commonaccent}, an open-source model trained to classify 16 English accents. We find that cloned voices are not only perceived as more \textit{``native''}, but also shift toward a specific set of dominant Anglophone varieties, including U.S. and U.K. English as well as Commonwealth-associated varieties such as Canadian, Australian, and New Zealand English. This pattern is most pronounced for \texttt{Coqui-XTTS}, for which all cloned recordings are classified as Inner Circle varieties (Figure~\ref{fig:homogenization_accent}).

The cadence of the speakers is also transformed and homogenized, with average clip duration being slightly shorter (source vs cloned, 5.11 to 4.92 seconds). We visualize the entropy of source and cloned recording durations and find the entropy of cloned durations to be lower (source vs cloned, 2.06 to 1.86) than that of the entropy of source durations, highlighting a concentration of audio durations, Figure \ref{fig:duration_entropy}.

\subsection{Cloned \textit{``Speakers''} Are Less Distinguishable}
In addition to audio recording analysis, we use classifier probes to assess how identifiable speakers are from their utterances. We represent every human and cloned clip as a 192-dimensional ECAPA-TDNN speaker embedding \citep{desplanques2020ecapa}. If voice cloning produces homogenization, then a classifier of fixed capacity should find it harder to distinguish between cloned speakers than between the source voices from which they were derived.

We build random forest and support vector machine (SVM) classifiers and train one set of classifiers on the source recordings and the other set of classifiers on the cloned recordings. Training data is sampled from speakers who have all nine valid utterances ($n=43$), and we randomly select five as training examples and hold out the remaining four for evaluation. Probes are fit for each cloning system and we report means over 20 random splits.

Speaker identity is recoverable from source recordings but less so from their ``cloned'' counterparts (Table \ref{tab:speaker_id}). A probe pooled over all three systems' clones drops from 94\% to 81\% (random forest) and from 99\% to 87\% (SVM); per system, the weakest is \texttt{ElevenLabs V3} at 61\% and 72\% (Table \ref{tab:speaker_id}). Identification errors also scatter more widely: the mean number of distinct incorrect classifications per speaker rises from 0.23 on sources to 1.44 for \texttt{ElevenLabs V3}; suggesting that misidentification is spread across a broader set of identities. The margin between same-speaker and different-speaker cosine similarity also narrows from 0.378 on source embeddings to 0.232 on \texttt{ElevenLabs} cloned embeddings, Appendix Table \ref{tab:verification}.

\begin{table}[h]
\centering
\caption{Speaker-identity probe on ECAPA-TDNN embeddings, per cloning system. \emph{Spread} is the mean number of distinct incorrect speakers a true speaker's clips are assigned to, $n=43$.}
\label{tab:speaker_id}
\small
\begin{tabular}{lcccc}
\toprule
\textbf{Metric} & \textbf{Source} & \textbf{\texttt{ChatterBox}} & \textbf{\texttt{Coqui}} & \textbf{\texttt{ElevenLabs}} \\
\midrule
Top-1 accuracy (Random Forest) & 94\% & 79\% & 81\% & 61\% \\
Top-1 accuracy (SVM) & 99\% & 93\% & 93\% & 72\% \\
\addlinespace
Mean incorrect-speaker spread & 0.23 & 0.79 & 0.71 & 1.44 \\
\bottomrule
\end{tabular}
\end{table}

\subsection{Audio Embeddings Eventually Converge}
Beyond the confusion between speaker identities observed in our classification probe, we find evidence that voice cloning applies a directional style transfer that will eventually converge across iterative rounds, further evidence of homogenization.

\begin{figure}[h!]
\begin{subfigure}[t]{\textwidth}
\centering
\includegraphics[width=0.82\linewidth]{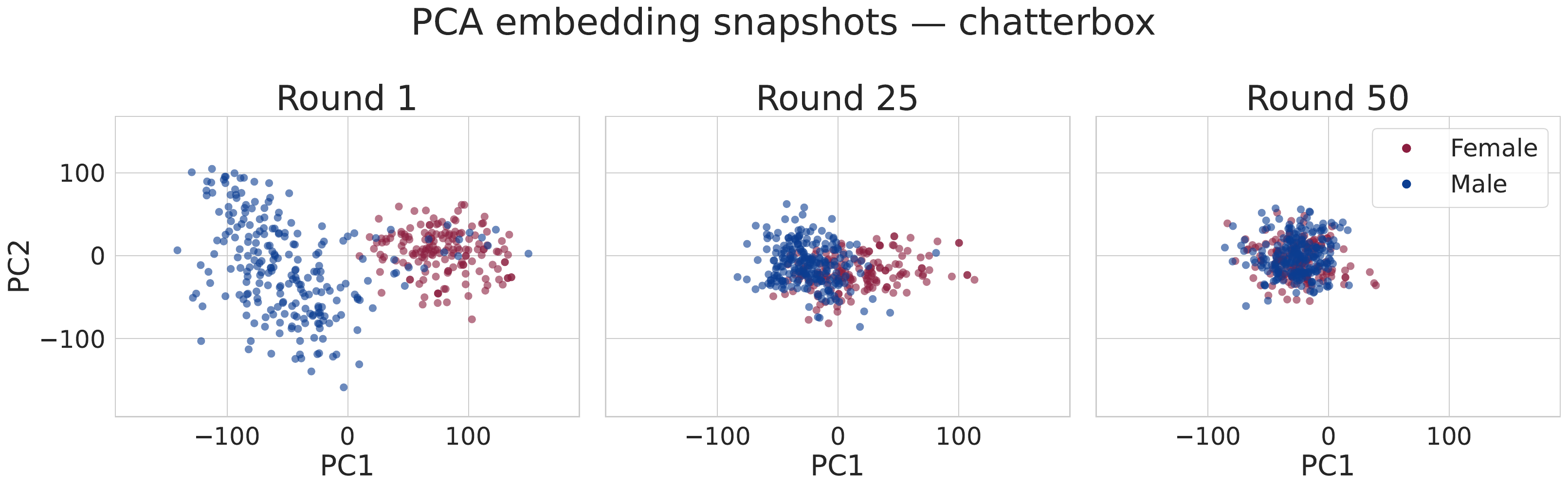}
\label{fig:all-three}
\end{subfigure}
\begin{subfigure}[b]{\textwidth}
\centering
\includegraphics[width=.52\textwidth]{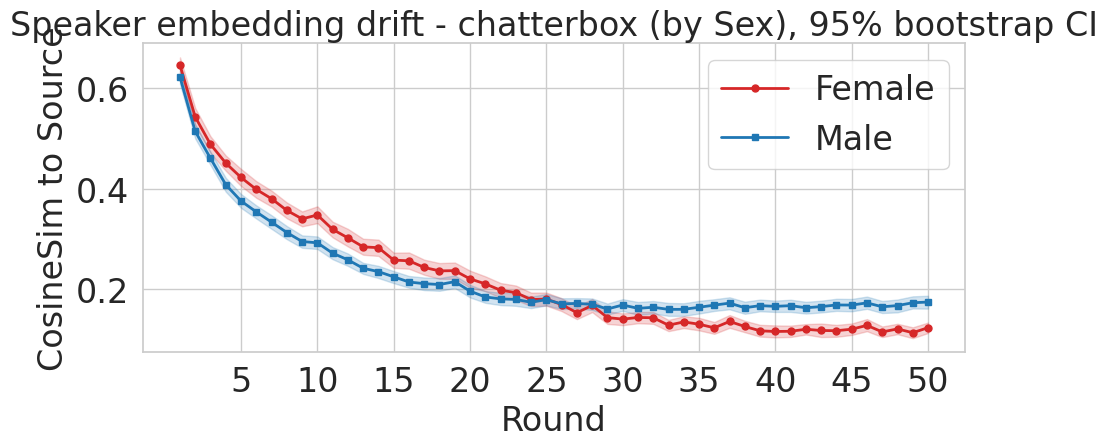}
\end{subfigure}
\hfill
\begin{subfigure}[b]{\textwidth}
\centering
\includegraphics[width=.48\textwidth]{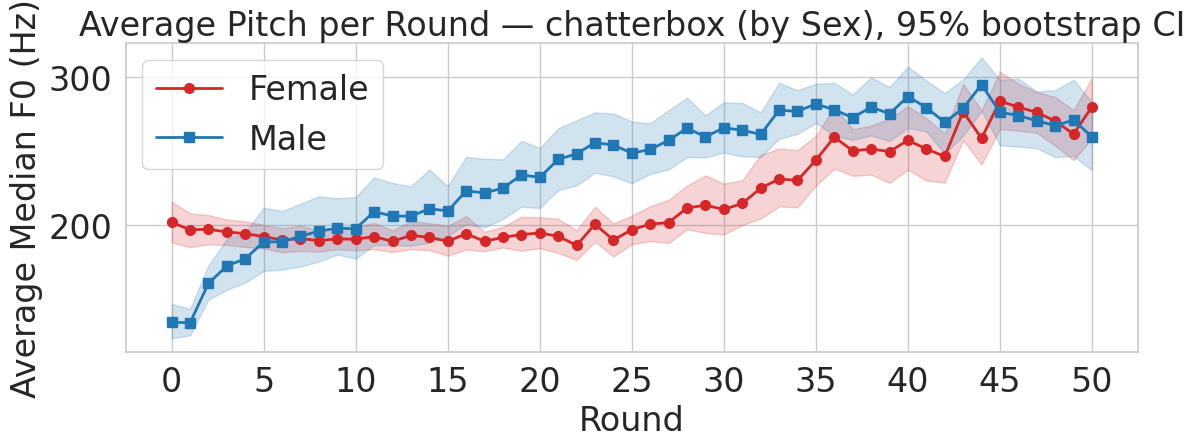}
\end{subfigure}
\caption{Changes to cloned audio across 50 rounds of repeated cloning with \texttt{Chatterbox}. (Top) Audio embeddings via PCA. (Middle) Cosine similarity to source. (Bottom) Pitch change from source; both show 95\% bootstrap CIs of the mean.}
\label{fig:changes}
\end{figure}

True voice cloning, in the strict sense, would produce a copy indistinguishable from the reference — and cloning that copy would yield the same result again. Some noise is inevitable in practice, since our reference and target texts differ, but if that noise were random, we would expect the mean of cloned embeddings to remain close to the source, even as individual clips drifted in embedding space. In our iterative cloning experiment, we take participants who have 9 valid sentences $(n=43)$ and repeatedly clone their voices using the cross-sentencing cloning method for 50 rounds. We then embed the cloned audio over time and track their movement, and we see that the transformation is systematic, directional, and convergent. After fifty rounds, embeddings cluster significantly closer together (with the radii of the approximate bounding sphere going from 366 to 336 in Euclidean distance, Figure~\ref{fig:changes}).\footnote{Calculated using the distance between the centroid embedding and the farthest individual speaker embedding.} 

We further characterize this drift through cosine similarity, pitch analysis, and emotion classification using \texttt{NVIDIA's Audio2Emotion-v3.0} model \citep{nvidia2025audio2face3d}. We see a significant drop in cosine similarity from the source embedding across the rounds; we see a dramatic increase in pitches for both male and female speakers, Figure \ref{fig:changes}, and pronounced changes to the emotions classified with a significant increase in predictions of anger (potentially due to increasingly empathic reading of the passage, Figure \ref{fig:emotion_change}).\footnote{Unlike \S\ref{sec:findings1}, here, machine classification is used instead of humans to produce labels after 50 rounds of cloning.} While users are unlikely to clone the same voice fifty times in practice, this experiment serves as an illustrative example of the directional and convergent nature of the voice transformation.

\section{Discussion of Potential Harms}
\label{sec:harms}
Our findings show discernible differences in voice characteristics between human recordings and their cloned counterparts. Because we already mitigate key confounds through manual quality filtering, including human-read disfluencies and lower microphone quality, these differences are unlikely to be artifacts of recording quality or delivery. In addition, annotations are done in pairs --- the same annotator rates the source and its clone --- so cross-annotator differences do not drive these comparisons. What we report are therefore paired, within-annotator differences between source recordings and their cloned counterparts. Here, we discuss the potential limitations and harms of our findings.

\paragraph{Unfaithful Voice Cloning} Voice cloning technology demonstrates limitations to faithfully clone voices. Instead, it applies systematic style transfers (desirable or not) to voices. This transformation has not been made explicit in these systems, and it is unclear if it is possible to perform voice cloning without these stylistic changes. The homogenizing effect of this style transfer over iterative cloning additionally puts the speech model at risk of modal collapse \citep{shumailov2024ai, ICLR2024_ebc042e7}, as synthetic audio data is often used to train and finetune models \citep{cornell2024generating, zhao2024advancing}. 

\paragraph{Behavioral and institutional risk at scale} A potential risk of this systematic application of style transfer is the downstream impacts it poses to human behavior. As text-to-speech and voice cloning gain popularity, voices that sound more persuasive or trustworthy raise safety concerns. Widespread deployment could erode human agency and increase disclosure of sensitive information, and could make misuse even worse. For example, an insurance company whose synthetic voice sounds more persuasive to users disputing a claim, or a fraudulent caller who extracts more personal information from a target. These risks are heightened by the accessibility of current open-sourced or cheap zero-shot voice-cloning systems and by the small amount of reference data (as little as a few seconds) required to generate a convincing clone. Greater transparency about how generated voices are produced and labeled \citep{chi_ea_transparent_ai_disclosures} should be investigated in the scientific community and in the public policy sphere.

\paragraph{Normative risk: Whose voice counts as human?} This failure of voice cloning fidelity matters not only for individual users who expect their voices to be preserved, but also for broader questions of labor and representation. When cloned voices are deployed in place of human speakers, they may not merely substitute for vocal labor, but reshape it according to narrow norms of what a voice should sound like. Voice cloning could contribute to reinforcing a narrow picture of what counts as a \textit{``human-like''} voice. Cloned voices were rated as more human-like than their source recordings--- a speech analogue of hyperrealism in generative media \citep{hyperrealism-ai-miller}. We hypothesize that listeners’ prototype of a \textit{``humanlike''} voice might value fluency and varieties of \textit{``Standard''} English, making clones seem more humanlike than source recordings.  Voice cloning may therefore encode and amplify existing language ideologies by making cloned voices sound more authoritative, warmer, more native-like, or even more "human" than the originals. Without clear watermarkings of generated speech, continued use of synthetic voices in downstream applications could further entrench the belief that to sound humanlike is to be fluent and to use standard English accents. 

\section{Conclusion}
Voice cloning is often framed as a technology for faithfully preserving or replicating a person’s voice. However, our findings show that voice cloning systems are actually applying style transformations on source voices in a homogenizing way. These transformations alter the way speakers are perceived, reduce vocal distinctiveness, and privilege standard accents. As speech technologies continue to grow in adoption, it is critical for researchers and policymakers to have a clear and empirical understanding of how these systems affect the voices they claim to preserve.

\paragraph{Scope and limitations.}
Our study is limited as we are analyzing zero-shot cloning of read English speech, recorded by U.S.-based speakers who are mostly non-native English speakers, across three systems and a single standard passage. We do not test spontaneous or conversational speech, non-English targets, or few-shot and finetuned cloning. As mentioned in the main text, the trust and disclosure outcomes are self-reported perceptions rather than observed behavior; establishing behavioral consequences requires in-situ study with incentivized users. Speaker sex is the binary category supplied by the recruitment platform and does not capture gender identity. Finally, our annotators are monolingual English speakers, so the perceptual shifts we report may not generalize to a broader, non-English and/or multilingual group.

\begin{ack}
Many thanks to all the online workers who participated in our studies! Funding provided by TogetherAI.
\end{ack}

\bibliography{bibliography}

@misc{le2023voiceboxtextguidedmultilingualuniversal,
   title={Voicebox: Text-Guided Multilingual Universal Speech Generation at Scale}, 
   author={Matthew Le and Apoorv Vyas and Bowen Shi and Brian Karrer and Leda Sari and Rashel Moritz and Mary Williamson and Vimal Manohar and Yossi Adi and Jay Mahadeokar and Wei-Ning Hsu},
   year={2023},
   eprint={2306.15687},
   archivePrefix={arXiv},
   primaryClass={eess.AS},
   url={https://arxiv.org/abs/2306.15687}, 
}

@inproceedings{arik2018neuralvoicecloningsamples,
   title={Neural Voice Cloning with a Few Samples}, 
   author={Sercan O. Arik and Jitong Chen and Kainan Peng and Wei Ping and Yanqi Zhou},
   year={2018},
   eprint={1802.06006},
   booktitle={Advances in Neural Information Processing Systems},
   primaryClass={cs.CL},
   url={https://arxiv.org/abs/1802.06006}, 
}

@misc{ju2024naturalspeech3zeroshotspeech,
   title={NaturalSpeech 3: Zero-Shot Speech Synthesis with Factorized Codec and Diffusion Models}, 
   author={Zeqian Ju and Yuancheng Wang and Kai Shen and Xu Tan and Detai Xin and Dongchao Yang and Yanqing Liu and Yichong Leng and Kaitao Song and Siliang Tang and Zhizheng Wu and Tao Qin and Xiang-Yang Li and Wei Ye and Shikun Zhang and Jiang Bian and Lei He and Jinyu Li and Sheng Zhao},
   year={2024},
   eprint={2403.03100},
   archivePrefix={arXiv},
   primaryClass={eess.AS},
   url={https://arxiv.org/abs/2403.03100}, 
}

@misc{shen2023naturalspeech2latentdiffusion,
   title={NaturalSpeech 2: Latent Diffusion Models are Natural and Zero-Shot Speech and Singing Synthesizers}, 
   author={Kai Shen and Zeqian Ju and Xu Tan and Yanqing Liu and Yichong Leng and Lei He and Tao Qin and Sheng Zhao and Jiang Bian},
   year={2023},
   eprint={2304.09116},
   archivePrefix={arXiv},
   primaryClass={eess.AS},
   url={https://arxiv.org/abs/2304.09116}, 
}

@misc{wang2023neuralcodeclanguagemodels,
   title={Neural Codec Language Models are Zero-Shot Text to Speech Synthesizers}, 
   author={Chengyi Wang and Sanyuan Chen and Yu Wu and Ziqiang Zhang and Long Zhou and Shujie Liu and Zhuo Chen and Yanqing Liu and Huaming Wang and Jinyu Li and Lei He and Sheng Zhao and Furu Wei},
   year={2023},
   eprint={2301.02111},
   archivePrefix={arXiv},
   primaryClass={cs.CL},
   url={https://arxiv.org/abs/2301.02111}, 
}

@inproceedings{ji-etal-2025-controlspeech,
 title = "{C}ontrol{S}peech: Towards Simultaneous and Independent Zero-shot Speaker Cloning and Zero-shot Language Style Control",
 author = "Ji, Shengpeng  and
   Chen, Qian  and
   Wang, Wen  and
   Zuo, Jialong  and
   Fang, Minghui  and
   Jiang, Ziyue  and
   Huang, Hai  and
   Wang, Zehan  and
   Cheng, Xize  and
   Zheng, Siqi  and
   Zhao, Zhou",
 editor = "Che, Wanxiang  and
   Nabende, Joyce  and
   Shutova, Ekaterina  and
   Pilehvar, Mohammad Taher",
 booktitle = "Proceedings of the 63rd Annual Meeting of the Association for Computational Linguistics (Volume 1: Long Papers)",
 month = jul,
 year = "2025",
 address = "Vienna, Austria",
 publisher = "Association for Computational Linguistics",
 url = "https://aclanthology.org/2025.acl-long.346/",
 doi = "10.18653/v1/2025.acl-long.346",
 pages = "6966--6981",
 ISBN = "979-8-89176-251-0"
}

@misc{chen2024emoknobenhancevoicecloning,
   title={EmoKnob: Enhance Voice Cloning with Fine-Grained Emotion Control},
   author={Haozhe Chen and Run Chen and Julia Hirschberg},
   year={2024},
   eprint={2410.00316},
   archivePrefix={arXiv},
   primaryClass={cs.CL},
   url={https://arxiv.org/abs/2410.00316},
}

@INPROCEEDINGS{promptts,
  author={Guo, Zhifang and Leng, Yichong and Wu, Yihan and Zhao, Sheng and Tan, Xu},
  booktitle={ICASSP 2023 - 2023 IEEE International Conference on Acoustics, Speech and Signal Processing (ICASSP)}, 
  title={Prompttts: Controllable Text-To-Speech With Text Descriptions}, 
  year={2023},
  volume={},
  number={},
  pages={1-5},
  doi={10.1109/ICASSP49357.2023.10096285}}

@misc{yang2023instructttsmodellingexpressivetts,
   title={InstructTTS: Modelling Expressive TTS in Discrete Latent Space with Natural Language Style Prompt}, 
   author={Dongchao Yang and Songxiang Liu and Rongjie Huang and Chao Weng and Helen Meng},
   year={2023},
   eprint={2301.13662},
   archivePrefix={arXiv},
   primaryClass={cs.SD},
   url={https://arxiv.org/abs/2301.13662}, 
}

@misc{qin2024openvoiceversatileinstantvoice,
   title={OpenVoice: Versatile Instant Voice Cloning}, 
   author={Zengyi Qin and Wenliang Zhao and Xumin Yu and Xin Sun},
   year={2024},
   eprint={2312.01479},
   archivePrefix={arXiv},
   primaryClass={cs.SD},
   url={https://arxiv.org/abs/2312.01479}, 
}

@misc{anastassiou2024seedttsfamilyhighqualityversatile,
   title={Seed-TTS: A Family of High-Quality Versatile Speech Generation Models}, 
   author={Philip Anastassiou and Jiawei Chen and Jitong Chen and Yuanzhe Chen and Zhuo Chen and Ziyi Chen and Jian Cong and Lelai Deng and Chuang Ding and Lu Gao and Mingqing Gong and Peisong Huang and Qingqing Huang and Zhiying Huang and Yuanyuan Huo and Dongya Jia and Chumin Li and Feiya Li and Hui Li and Jiaxin Li and Xiaoyang Li and Xingxing Li and Lin Liu and Shouda Liu and Sichao Liu and Xudong Liu and Yuchen Liu and Zhengxi Liu and Lu Lu and Junjie Pan and Xin Wang and Yuping Wang and Yuxuan Wang and Zhen Wei and Jian Wu and Chao Yao and Yifeng Yang and Yuanhao Yi and Junteng Zhang and Qidi Zhang and Shuo Zhang and Wenjie Zhang and Yang Zhang and Zilin Zhao and Dejian Zhong and Xiaobin Zhuang},
   year={2024},
   eprint={2406.02430},
   archivePrefix={arXiv},
   primaryClass={eess.AS},
   url={https://arxiv.org/abs/2406.02430}, 
}

@inproceedings{li-etal-2024-pause,
 title = "Pause-Aware Automatic Dubbing using {LLM} and Voice Cloning",
 author = "Li, Yuang  and
   Guo, Jiaxin  and
   Zhang, Min  and
   Miaomiao, Ma  and
   Rao, Zhiqiang  and
   Zhang, Weidong  and
   He, Xianghui  and
   Wei, Daimeng  and
   Yang, Hao",
 editor = "Salesky, Elizabeth  and
   Federico, Marcello  and
   Carpuat, Marine",
 booktitle = "Proceedings of the 21st International Conference on Spoken Language Translation (IWSLT 2024)",
 month = aug,
 year = "2024",
 address = "Bangkok, Thailand (in-person and online)",
 publisher = "Association for Computational Linguistics",
 url = "https://aclanthology.org/2024.iwslt-1.2/",
 doi = "10.18653/v1/2024.iwslt-1.2",
 pages = "12--16",
}

@article{wells2022s,
  title={What's in a voice? The legal implications of voice cloning},
  author={Wells-Edwards, Bryn},
  journal={Ariz. L. Rev.},
  volume={64},
  pages={1213},
  year={2022},
  publisher={HeinOnline}
}

@inproceedings{platnick-etal-2024-preset,
 title = "Preset-Voice Matching for Privacy Regulated Speech-to-Speech Translation Systems",
 author = "Platnick, Daniel  and
   Abdelnour, Bishoy  and
   Earl, Eamon  and
   Kumar, Rahul  and
   Rezaei, Zahra  and
   Tsangaris, Thomas  and
   Lagum, Faraj",
 editor = "Habernal, Ivan  and
   Ghanavati, Sepideh  and
   Ravichander, Abhilasha  and
   Jain, Vijayanta  and
   Thaine, Patricia  and
   Igamberdiev, Timour  and
   Mireshghallah, Niloofar  and
   Feyisetan, Oluwaseyi",
 booktitle = "Proceedings of the Fifth Workshop on Privacy in Natural Language Processing",
 month = aug,
 year = "2024",
 address = "Bangkok, Thailand",
 publisher = "Association for Computational Linguistics",
 url = "https://aclanthology.org/2024.privatenlp-1.6/",
 pages = "52--62",
}

@book{van1972speech,
  title={Speech correction},
  author={Van Riper, Charles},
  year={1972},
  publisher={Prentice-Hall New York}
}

@inproceedings{zhou-etal-2025-rel,
 title = "{REL}-{A}.{I}.: An Interaction-Centered Approach To Measuring Human-{LM} Reliance",
 author = "Zhou, Kaitlyn  and
   Hwang, Jena D.  and
   Ren, Xiang  and
   Dziri, Nouha  and
   Jurafsky, Dan  and
   Sap, Maarten",
 editor = "Chiruzzo, Luis  and
   Ritter, Alan  and
   Wang, Lu",
 booktitle = "Proceedings of the 2025 Conference of the Nations of the Americas Chapter of the Association for Computational Linguistics: Human Language Technologies (Volume 1: Long Papers)",
 month = apr,
 year = "2025",
 address = "Albuquerque, New Mexico",
 publisher = "Association for Computational Linguistics",
 url = "https://aclanthology.org/2025.naacl-long.556/",
 doi = "10.18653/v1/2025.naacl-long.556",
 pages = "11148--11167",
 ISBN = "979-8-89176-189-6",
}

@article{hyperrealism-ai-miller,
author = {Elizabeth J. Miller and Ben A. Steward and Zak Witkower and Clare A. M. Sutherland and Eva G. Krumhuber and Amy Dawel},
title ={AI Hyperrealism: Why AI Faces Are Perceived as More Real Than Human Ones},

journal = {Psychological Science},
volume = {34},
number = {12},
pages = {1390-1403},
year = {2023},
doi = {10.1177/09567976231207095},
note ={PMID: 37955384},
URL = {https://doi.org/10.1177/09567976231207095},
eprint = {https://doi.org/10.1177/09567976231207095}}

@article{nightingale,
author = {Sophie J. Nightingale  and Hany Farid },
title = {AI-synthesized faces are indistinguishable from real faces and more trustworthy},
journal = {Proceedings of the National Academy of Sciences},
volume = {119},
number = {8},
pages = {e2120481119},
year = {2022},
doi = {10.1073/pnas.2120481119},
URL = {https://www.pnas.org/doi/abs/10.1073/pnas.2120481119},
eprint = {https://www.pnas.org/doi/pdf/10.1073/pnas.2120481119}}

@inproceedings{accent_bias_facct,
author = {Michel, Shira and Kaur, Sufi and Gillespie, Sarah Elizabeth and Gleason, Jeffrey and Wilson, Christo and Ghosh, Avijit},
title = {“It’s not a representation of me”: Examining Accent Bias and Digital Exclusion in Synthetic AI Voice Services},
year = {2025},
isbn = {9798400714825},
publisher = {Association for Computing Machinery},
address = {New York, NY, USA},
url = {https://doi.org/10.1145/3715275.3732018},
doi = {10.1145/3715275.3732018},
booktitle = {Proceedings of the 2025 ACM Conference on Fairness, Accountability, and Transparency},
pages = {228–245},
numpages = {18},
location = {
},
series = {FAccT '25}
}

@inproceedings{cscw_du_voice_cloning,
author = {Du, Jiachen and Huang, Hanyu and Zou, Xinkai and Yin, Shuzi and Gao, Bingjie and Fu, Xinyi},
title = {The Social Dynamics of Voice Cloning: Trust, Privacy, and Ethical Tensions When Sharing Your AI Voice Replica},
year = {2025},
isbn = {9798400714801},
publisher = {Association for Computing Machinery},
address = {New York, NY, USA},
url = {https://doi.org/10.1145/3715070.3749244},
doi = {10.1145/3715070.3749244},
booktitle = {Companion Publication of the 2025 Conference on Computer-Supported Cooperative Work and Social Computing},
pages = {307–311},
numpages = {5},
location = {
},
series = {CSCW Companion '25}
}

@inproceedings{chi_audio_less_trustworthy,
author = {Do, Tiffany D. and McMahan, Ryan P. and Wisniewski, Pamela J.},
title = {A New Uncanny Valley? The Effects of Speech Fidelity and Human Listener Gender on Social Perceptions of a Virtual-Human Speaker},
year = {2022},
isbn = {9781450391573},
publisher = {Association for Computing Machinery},
address = {New York, NY, USA},
url = {https://doi.org/10.1145/3491102.3517564},
doi = {10.1145/3491102.3517564},
booktitle = {Proceedings of the 2022 CHI Conference on Human Factors in Computing Systems},
articleno = {424},
numpages = {11},
location = {New Orleans, LA, USA},
series = {CHI '22}
}

@article{mai2023warning,
  title={Warning: Humans cannot reliably detect speech deepfakes},
  author={Mai, Kimberly T and Bray, Sergi and Davies, Toby and Griffin, Lewis D},
  journal={Plos one},
  volume={18},
  number={8},
  pages={e0285333},
  year={2023},
  publisher={Public Library of Science}
}

@inproceedings{chi_ea_transparent_ai_disclosures,
author = {El Ali, Abdallah and Venkatraj, Karthikeya Puttur and Morosoli, Sophie and Naudts, Laurens and Helberger, Natali and Cesar, Pablo},
title = {Transparent AI Disclosure Obligations: Who, What, When, Where, Why, How},
year = {2024},
isbn = {9798400703317},
publisher = {Association for Computing Machinery},
address = {New York, NY, USA},
url = {https://doi.org/10.1145/3613905.3650750},
doi = {10.1145/3613905.3650750},
booktitle = {Extended Abstracts of the CHI Conference on Human Factors in Computing Systems},
articleno = {342},
numpages = {11},
location = {Honolulu, HI, USA},
series = {CHI EA '24}
}

@article{lavan2025voice,
  title={Voice clones sound realistic but not (yet) hyperrealistic},
  author={Lavan, Nadine and Irvine, Mairi and Rosi, Victor and McGettigan, Carolyn},
  journal={PLoS One},
  volume={20},
  number={9},
  pages={e0332692},
  year={2025},
  publisher={Public Library of Science}
}

@inproceedings{chavan_chi_ea_aac,
author = {R Chavan, Durwa and Moon, Prachi and Dixon, Emma},
title = {Public Reflections on the Use of Augmentative and Alternative Communication (AAC) Devices by People with I/DD in Everyday Life},
year = {2025},
isbn = {9798400706769},
publisher = {Association for Computing Machinery},
address = {New York, NY, USA},
url = {https://doi.org/10.1145/3663547.3759720},
doi = {10.1145/3663547.3759720},
booktitle = {Proceedings of the 27th International ACM SIGACCESS Conference on Computers and Accessibility},
articleno = {161},
numpages = {4},
location = {
},
series = {ASSETS '25}
}

@article{fiske2007universal,
  title={Universal dimensions of social cognition: Warmth and competence},
  author={Fiske, Susan T and Cuddy, Amy JC and Glick, Peter},
  journal={Trends in cognitive sciences},
  volume={11},
  number={2},
  pages={77--83},
  year={2007},
  publisher={Elsevier}
}

@inproceedings{almeda2025labor,
  title={Labor, Power, and Belonging: The Work of Voice in the Age of AI Reproduction},
  author={Almeda, Shm and Netzorg, Robin and Li, Isabel and Tam, Ethan and Ma, Skyla and Wei, Bob Tianqi},
  booktitle={Proceedings of the 2025 ACM Conference on Fairness, Accountability, and Transparency},
  pages={1238--1249},
  year={2025}
}

@article{pataranutaporn2021ai,
  title={AI-generated characters for supporting personalized learning and well-being},
  author={Pataranutaporn, Pat and Danry, Valdemar and Leong, Joanne and Punpongsanon, Parinya and Novy, Dan and Maes, Pattie and Sra, Misha},
  journal={Nature Machine Intelligence},
  volume={3},
  number={12},
  pages={1013--1022},
  year={2021},
  publisher={Nature Publishing Group UK London}
}

@article{wairagkar2025instantaneous,
  title={An instantaneous voice-synthesis neuroprosthesis},
  author={Wairagkar, Maitreyee and Card, Nicholas S and Singer-Clark, Tyler and Hou, Xianda and Iacobacci, Carrina and Miller, Lee M and Hochberg, Leigh R and Brandman, David M and Stavisky, Sergey D},
  journal={Nature},
  volume={644},
  number={8075},
  pages={145--152},
  year={2025},
  publisher={Nature Publishing Group UK London}
}

@article{zuluaga2023commonaccent,
  title={Commonaccent: Exploring large acoustic pretrained models for accent classification based on common voice},
  author={Zuluaga-Gomez, Juan and Ahmed, Sara and Visockas, Danielius and Subakan, Cem},
  journal={Interspeech},
  year={2023}
}

@inproceedings{radford2023robust,
  title={Robust speech recognition via large-scale weak supervision},
  author={Radford, Alec and Kim, Jong Wook and Xu, Tao and Brockman, Greg and McLeavey, Christine and Sutskever, Ilya},
  booktitle={International conference on machine learning},
  pages={28492--28518},
  year={2023},
  organization={PMLR}
}

@misc{nvidia2025audio2face3d,
      title={Audio2Face-3D: Audio-driven Realistic Facial Animation For Digital Avatars},
      author={Chaeyeon Chung and Ilya Fedorov and Michael Huang and Aleksey Karmanov and Dmitry Korobchenko and Roger Ribera and Yeongho Seol},
      year={2025},
      eprint={2508.16401},
      archivePrefix={arXiv},
      primaryClass={cs.GR},
      url={https://arxiv.org/abs/2508.16401},
      note={Authors listed in alphabetical order}
}

@article{nytimes2026koreaAI,
  title   = {South Korea Uses AI to Help Seniors with Dementia},
  author  = {Sang-Hun, Choe},
  journal = {The New York Times},
  year    = {2026},
  month   = apr,
  day     = {28},
  url     = {https://www.nytimes.com/2026/04/28/world/asia/korea-ai-seniors-dementia.html}
}

@book{bijker1987social,
  title     = {The Social Construction of Technological Systems: New Directions in the Sociology and History of Technology},
  editor    = {Bijker, Wiebe E. and Hughes, Thomas P. and Pinch, Trevor J.},
  year      = {1987},
  publisher = {MIT Press},
  address   = {Cambridge, MA}
}

@incollection{winner2017artifacts,
  title={Do artifacts have politics?},
  author={Winner, Langdon},
  booktitle={Computer ethics},
  pages={177--192},
  year={2017},
  publisher={Routledge}
}

@inproceedings{not_by_voice_hutiri,
author = {Hutiri, Wiebke and Papakyriakopoulos, Orestis and Xiang, Alice},
title = {Not My Voice! A Taxonomy of Ethical and Safety Harms of Speech Generators},
year = {2024},
isbn = {9798400704505},
publisher = {Association for Computing Machinery},
address = {New York, NY, USA},
url = {https://doi.org/10.1145/3630106.3658911},
doi = {10.1145/3630106.3658911},
booktitle = {Proceedings of the 2024 ACM Conference on Fairness, Accountability, and Transparency},
pages = {359–376},
numpages = {18},
location = {Rio de Janeiro, Brazil},
series = {FAccT '24}
}

@article{lee2026vocal,
  title   = {Vocal Identity Under Siege by AI Voice Cloning Technologies},
  author  = {Lee, Jyh-An and Sun, Xuan},
  journal = {Singapore Journal of Legal Studies},
  year    = {2026},
  pages   = {46--76}
}

@article{berkowitz2025look,
  title={Look Who’s Talking: Voice cloning as tension point between identity and data},
  author={Berkowitz, Adam Eric and Sweeney, Miriam E},
  journal={Philosophy \& Technology},
  volume={38},
  number={4},
  pages={131},
  year={2025},
  publisher={Springer}
}

@article{berkowitz2026simulating,
  title={Simulating Voice and the Simulacra of Voice Clones},
  author={Berkowitz, Adam Eric and Sweeney, Miriam E},
  journal={Philosophy \& Technology},
  volume={39},
  number={1},
  pages={4},
  year={2026},
  publisher={Springer}
}

@article{leuenberger2025role,
  title={The role of the voice for identity and implications for voice cloning technology},
  author={Leuenberger, Muriel},
  journal={Philosophy \& Technology},
  volume={38},
  number={4},
  pages={175},
  year={2025},
  publisher={Springer}
}

@article{agnew2024sound,
  title={Sound check: Auditing audio datasets},
  author={Agnew, William and Barnett, Julia and Chu, Annie and Hong, Rachel and Feffer, Michael and Netzorg, Robin and Jiang, Harry H and Awumey, Ezra and Das, Sauvik},
  journal={AIES},
  year={2024}
}

@article{shumailov2024ai,
  title={AI models collapse when trained on recursively generated data},
  author={Shumailov, Ilia and Shumaylov, Zakhar and Zhao, Yiren and Papernot, Nicolas and Anderson, Ross and Gal, Yarin},
  journal={Nature},
  volume={631},
  number={8022},
  pages={755--759},
  year={2024},
  publisher={Nature Publishing Group UK London}
}

@inproceedings{ICLR2024_ebc042e7,
 author = {Alemohammad, Sina and Casco-Rodriguez, Josue and Luzi, Lorenzo and Humayun, Ahmed Imtiaz and Babaei, Hossein and LeJeune, Daniel and Siahkoohi, Ali and Baraniuk, Richard},
 booktitle = {International Conference on Learning Representations},
 pages = {53581--53608},
 title = {Self-Consuming Generative Models Go {MAD}},
 url = {https://proceedings.iclr.cc/paper_files/paper/2024/file/ebc042e767de551803ccfcc45e2454f5-Paper-Conference.pdf},
 volume = {2024},
 year = {2024}
}

@article{cornell2024generating,
  title={Generating data with text-to-speech and large-language models for conversational speech recognition},
  author={Cornell, Samuele and Darefsky, Jordan and Duan, Zhiyao and Watanabe, Shinji},
  journal={Synthetic Data’s Transformative Role in Foundational Speech Models},
  year={2024}
}

@article{zhao2024advancing,
  title={Advancing speech language models by scaling supervised fine-tuning with over 60,000 hours of synthetic speech dialogue data},
  author={Zhao, Shuaijiang and Guo, Tingwei and Xiang, Bajian and Wan, Tongtang and Niu, Qiang and Zou, Wei and Li, Xiangang},
  journal={arXiv preprint arXiv:2412.01078},
  year={2024}
}

@inproceedings{labied2022overview,
  title={An overview of automatic speech recognition preprocessing techniques},
  author={Labied, Maria and Belangour, Abdessamad and Banane, Mouad and Erraissi, Allae},
  booktitle={2022 international conference on decision aid sciences and applications (DASA)},
  pages={804--809},
  year={2022},
  organization={IEEE}
}

@article{keerio2009preprocessing,
  title={On preprocessing of speech signals},
  author={Keerio, Ayaz and Mitra, Bhargav Kumar and Birch, Philip and Young, Rupert and Chatwin, Chris},
  journal={International Journal of Signal Processing},
  volume={5},
  number={3},
  pages={216--222},
  year={2009}
}

@article{zheng2025learning,
  title={Learning through AI-clones: Enhancing self-perception and presentation performance},
  author={Zheng, Qingxiao and Chen, Zhuoer and Huang, Yun},
  journal={Computers in Human Behavior: Artificial Humans},
  volume={3},
  pages={100117},
  year={2025},
  publisher={Elsevier}
}

@inproceedings{mogi_chatbot_self_disclosure,
author = {Mogi, Yamato and Akahori, Wataru and Yamashita, Naomi},
title = {Exploring the Effects of Different Chatbot Voice Identities on Self-Disclosure},
year = {2026},
isbn = {9798400722783},
publisher = {Association for Computing Machinery},
address = {New York, NY, USA},
url = {https://doi.org/10.1145/3772318.3790546},
doi = {10.1145/3772318.3790546},
articleno = {51},
numpages = {19},
location = {
},
series = {CHI '26}
}

@inproceedings{park_esl_clone,
author = {Park, Minju and Lee, Seunghyun and Ma, Juhwan and Yoon, Dongwook},
title = {AI Twin: Enhancing ESL Speaking Practice through AI Self-Clones of a Better Me},
year = {2026},
isbn = {9798400722783},
publisher = {Association for Computing Machinery},
address = {New York, NY, USA},
url = {https://doi.org/10.1145/3772318.3790266},
doi = {10.1145/3772318.3790266},
booktitle = {Proceedings of the 2026 CHI Conference on Human Factors in Computing Systems},
articleno = {78},
numpages = {21},
location = {
},
series = {CHI '26}
}

@article{titze1989physiologic,
  title={Physiologic and acoustic differences between male and female voices},
  author={Titze, Ingo R},
  journal={The Journal of the Acoustical Society of America},
  volume={85},
  number={4},
  pages={1699--1707},
  year={1989},
  publisher={Acoustical Society of America}
}

@article{smith2007discrimination,
  title={Discrimination of speaker sex and size when glottal-pulse rate and vocal-tract length are controlled},
  author={Smith, David RR and Walters, Thomas C and Patterson, Roy D},
  journal={The Journal of the Acoustical Society of America},
  volume={122},
  number={6},
  pages={3628--3639},
  year={2007},
  publisher={AIP Publishing}
}

@article{desplanques2020ecapa,
  title={{ECAPA-TDNN}: Emphasized channel attention, propagation and aggregation in tdnn based speaker verification},
  author={Desplanques, Brecht and Thienpondt, Jenthe and Demuynck, Kris},
  journal={Interspeech},
  year={2020}
}

@misc{chatterboxtts2025,
  author       = {{Resemble AI}},
  title        = {{Chatterbox-TTS}},
  year         = {2025},
  howpublished = {\url{https://github.com/resemble-ai/chatterbox}},
  note         = {GitHub repository}
}

@software{Eren_Coqui_TTS_2021,
author = {Eren, Gölge and {The Coqui TTS Team}},
doi = {10.5281/zenodo.6334862},
license = {MPL-2.0},
month = jan,
title = {{Coqui TTS}},
url = {https://github.com/coqui-ai/TTS},
version = {1.4},
year = {2021}
}

@misc{elevenlabs_v3_2026,
  author       = {{ElevenLabs}},
  title        = {Eleven v3},
  year         = {2026},
  howpublished = {\url{https://elevenlabs.io}},
  note         = {General availability release. Accessed: 2026-04}
}
\bibliographystyle{plainnat}

\appendix
\section{Methods Details}
\label{section:methods_details}

\clearpage
\subsection{Grandfather Passage}
\label{section:grandfather_passage}
\texttt{You wish to know about my grandfather. Well, he is nearly 93 years old, yet he still thinks as swiftly as ever. He dresses himself in an old black frock coat, usually several buttons missing. A long beard clings to his chin, giving those who observe him a pronounced feeling of the utmost respect. When he speaks, his voice is just a bit cracked and quivers a bit. Twice each day he plays skillfully and with zest upon a small organ. Except in the winter when the snow or ice prevents, he slowly takes a short walk in the open air each day. We have often urged him to walk more and smoke less, but he always answers, “Banana oil!” Grandfather likes to be modern in his language.}

\clearpage
\subsection{Privacy Terms}
\label{sec:privacy_terms}
The participants are asked to give their consent to use their audio data in voice cloning technology for research purposes only, and will be made aware that we will be: Opting out of proprietary model training. Asking for their data to be deleted from companies with proprietary models at the conclusion of this project.

For our research purposes, their data was also used in voice cloning technology on open-sourced models, which we will host locally (i.e., their data stays with our research team). For downstream use, their audio data may be used to perform analysis on how audio data and voice cloned data differ. presented to other online workers to gather annotations about the perception of artificial vs natural voice. Used to train models to recognize the difference between artificial and natural human speech. Shared anonymously online via a public research dataset that cannot be used for commercial purposes (explicit guidelines below).

Forbidden Uses of the public dataset include:
\begin{itemize}
    \item Generating, enabling, or promoting hate speech, harassment,
discrimination, misinformation, or culturally offensive or harmful content
    \item Beyond explicit research purposes, voice cloning, speaker impersonation,
or the creation of synthetic voices intended to resemble or replicate any
participant
    \item Attempting to identify, re-identify, or infer the identity of any participant,
including attempts to extract personally identifiable information from the audio or
associated metadata
    \item Any commercial, for-profit, or revenue-generating use, including product
development, advertising, or monetized services
    \item Any use that misrepresents, stereotypes, or falsely attributes
characteristics, language abilities, accents, or identities to the speakers
    \item Redistribution of the dataset under terms that conflict with or weaken these
restrictions
\end{itemize}

\begin{figure}
    \centering
    \includegraphics[width=\linewidth]{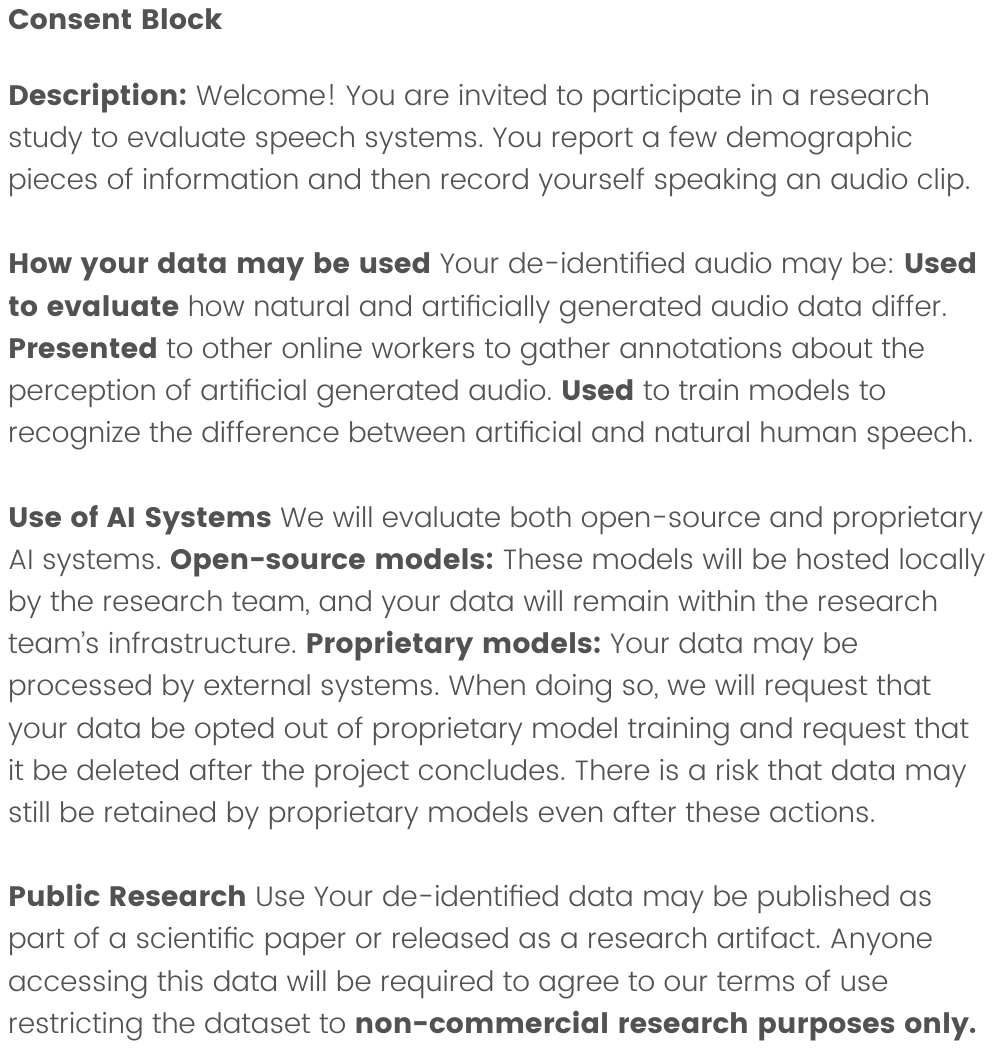}
    \caption{Screenshot of the consent block provided to the speaker.}
    \label{fig:placeholder}
\end{figure}

\begin{figure}
    \centering
    \includegraphics[width=.75\linewidth]{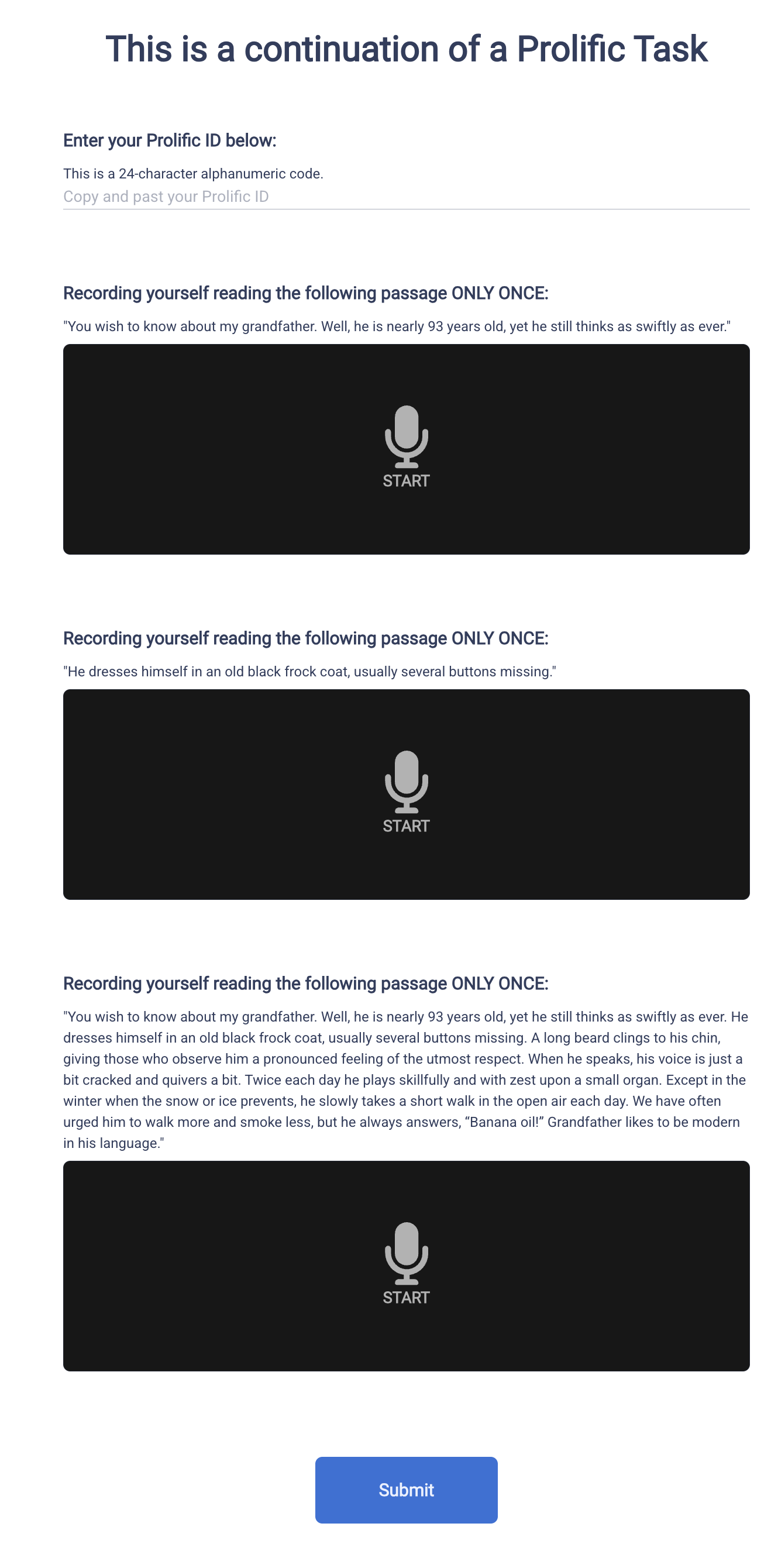}
    \caption{Screenshot of speaker task.}
    \label{fig:placeholder}
\end{figure}
\pagebreak

\begin{figure}
    \centering
    \includegraphics[width=0.95\linewidth]{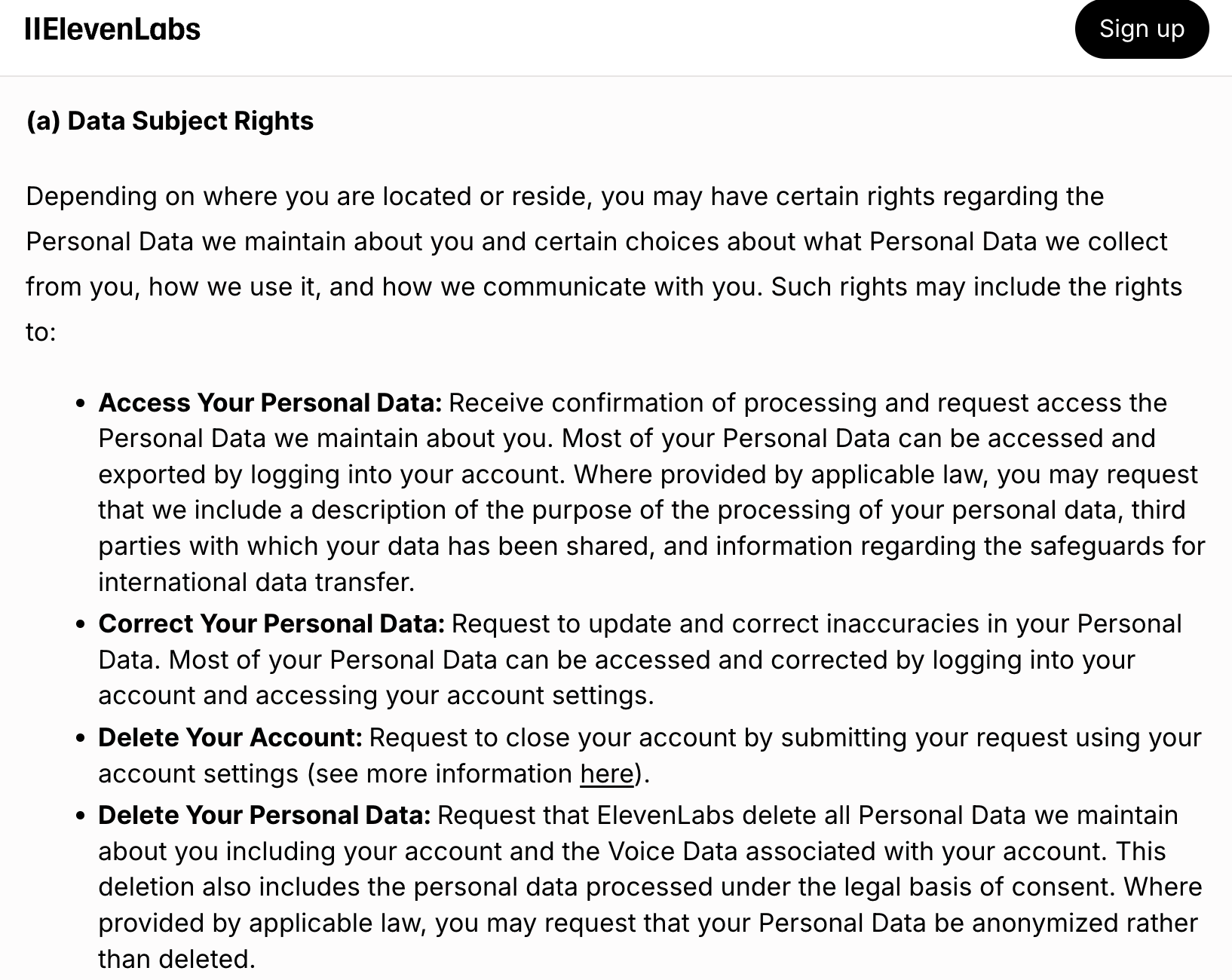}
    \caption{ElevenLabs Privacy Terms}
    \label{fig:placeholder}
\end{figure}

\clearpage
\subsection{Findings Ablations}
\begin{figure}
    \centering
    \includegraphics[width=1\linewidth]{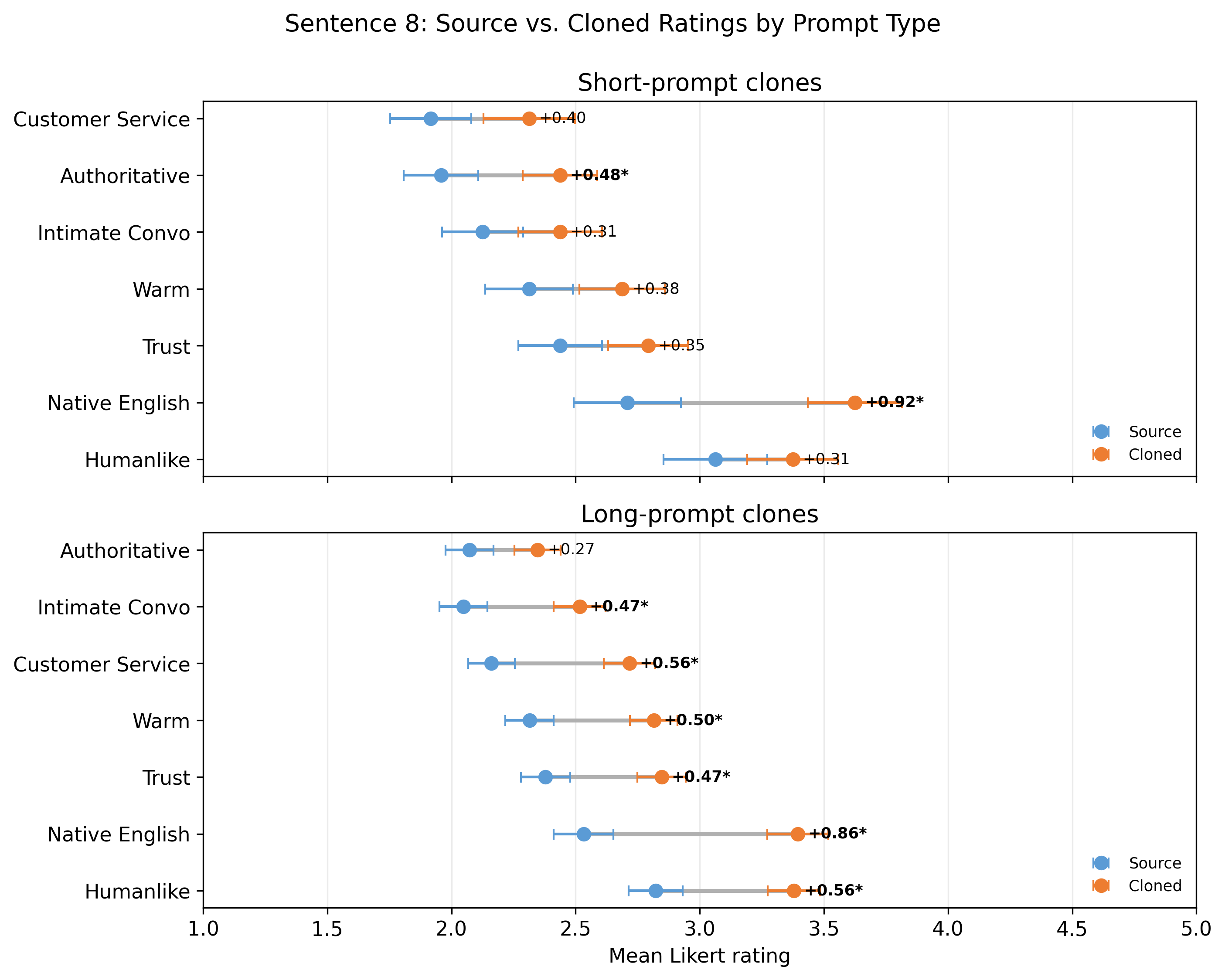}
    \caption{Comparison cloning with long vs short source clips (37 seconds versus 5 seconds). Long source clips are a concatenation of clips 1 through 7, whereas short clips are just clip 7.}
    \label{fig:long_vs_short}
\end{figure}

\begin{figure}
    \centering
    \includegraphics[width=1\linewidth]{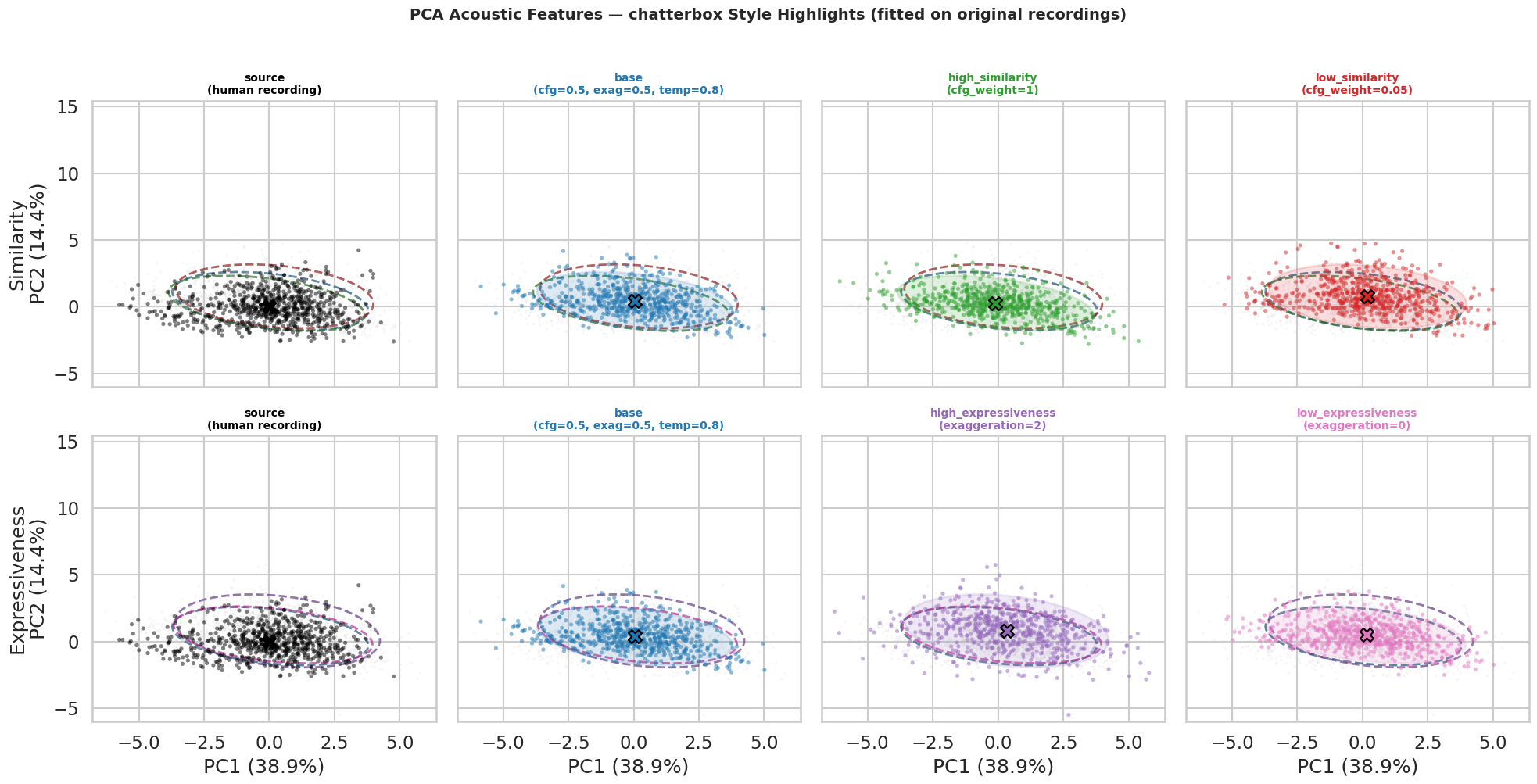}
    \caption{ PCA projections of Chatterbox acoustic embeddings under different styles. Across settings, generated clips remain close to the source distribution in embedding space, suggesting that the observed style shifts are not primarily explained by reduced speaker similarity or increased expressiveness.}
    \label{fig:embedding_styles}
\end{figure}

\begin{figure}
    \centering
    \includegraphics[width=1\linewidth]{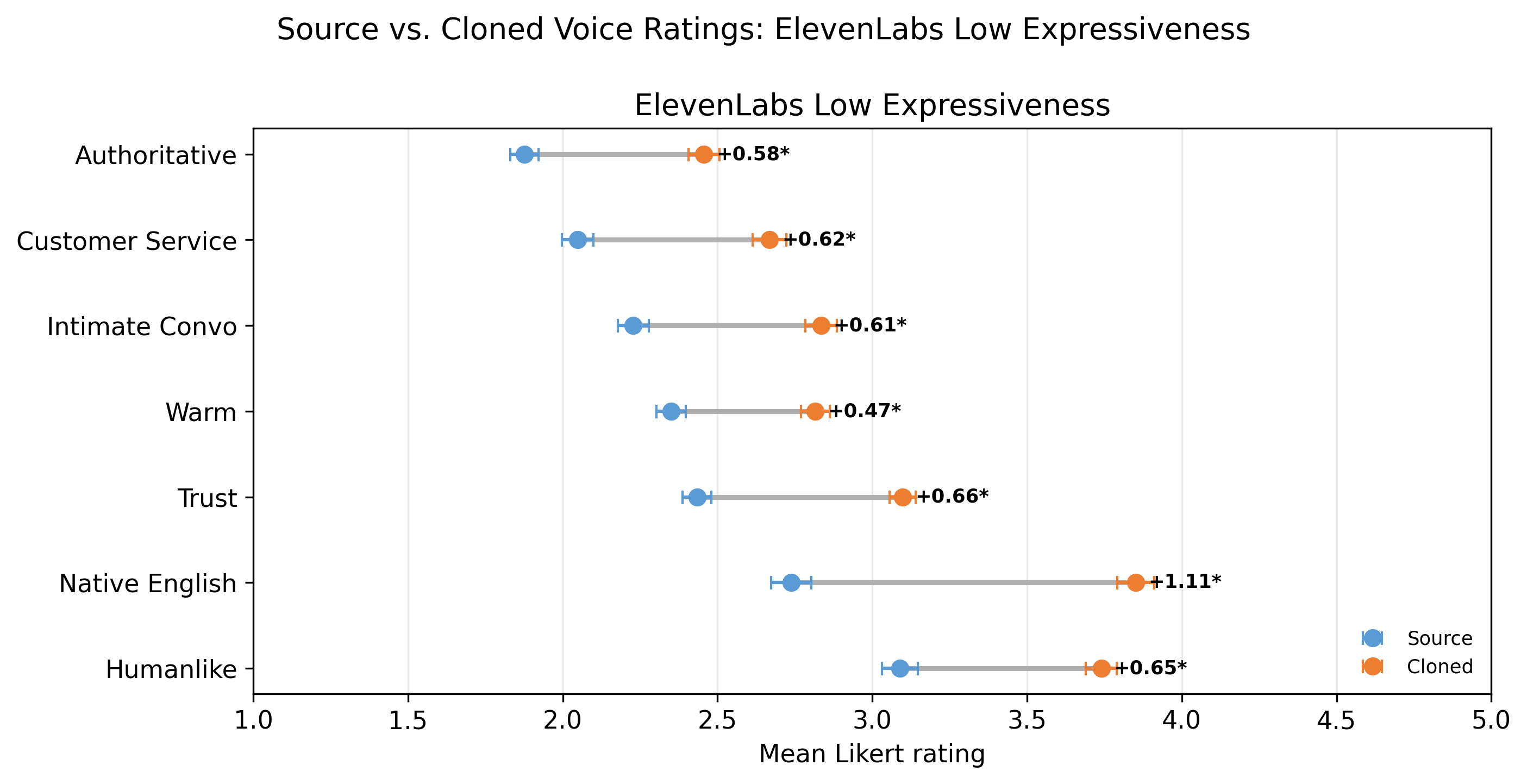}
    \caption{Human annotations on ElevenLabs clones under "low expressiveness"}
    \label{fig:elevenlabs_low_expressiveness}
\end{figure}

\begin{figure}
    \centering
    \includegraphics[width=0.95\linewidth]{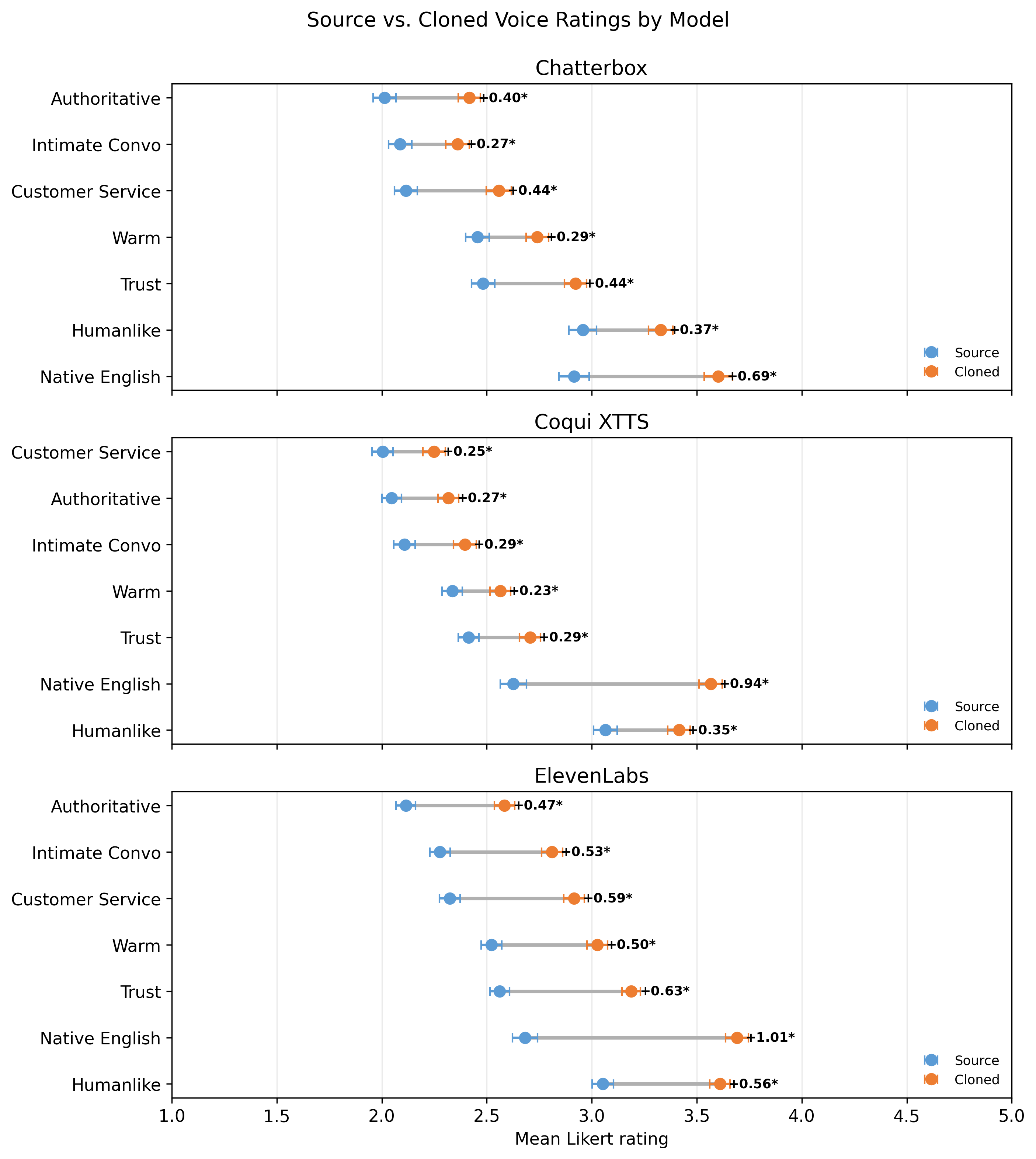}
    \caption{Rating differences between cloned and source voices by model.}
    \label{fig:ratings_by_model}
\end{figure}

\clearpage
\subsection{Context Length}
\begin{figure}
    \centering
    \includegraphics[width=0.95\linewidth]{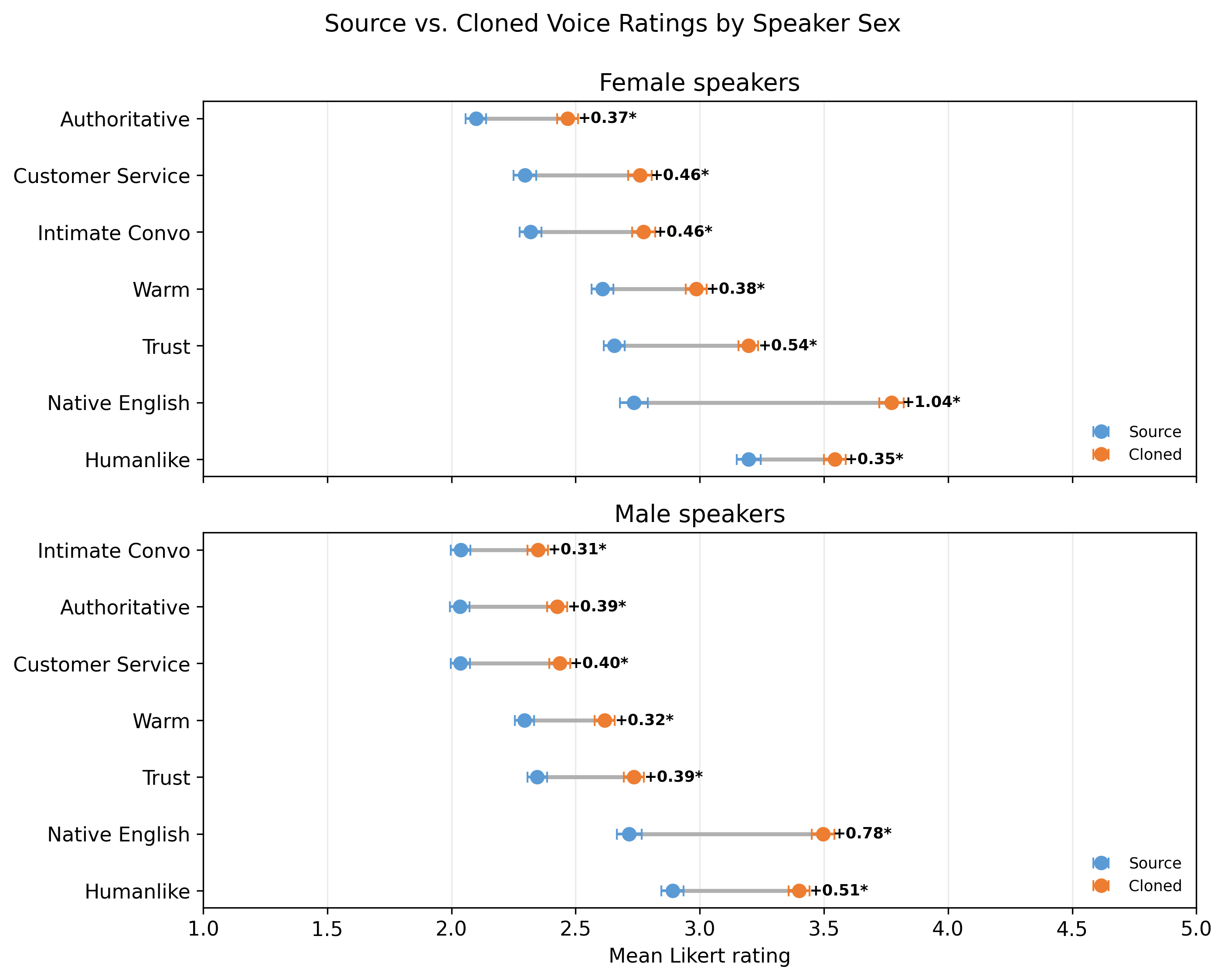}
    \caption{Rating differences between cloned and source voices by speaker sex.}
    \label{fig:ratings_by_speaker_sex}
\end{figure}

\begin{figure}
    \centering
    \includegraphics[width=0.65\linewidth]{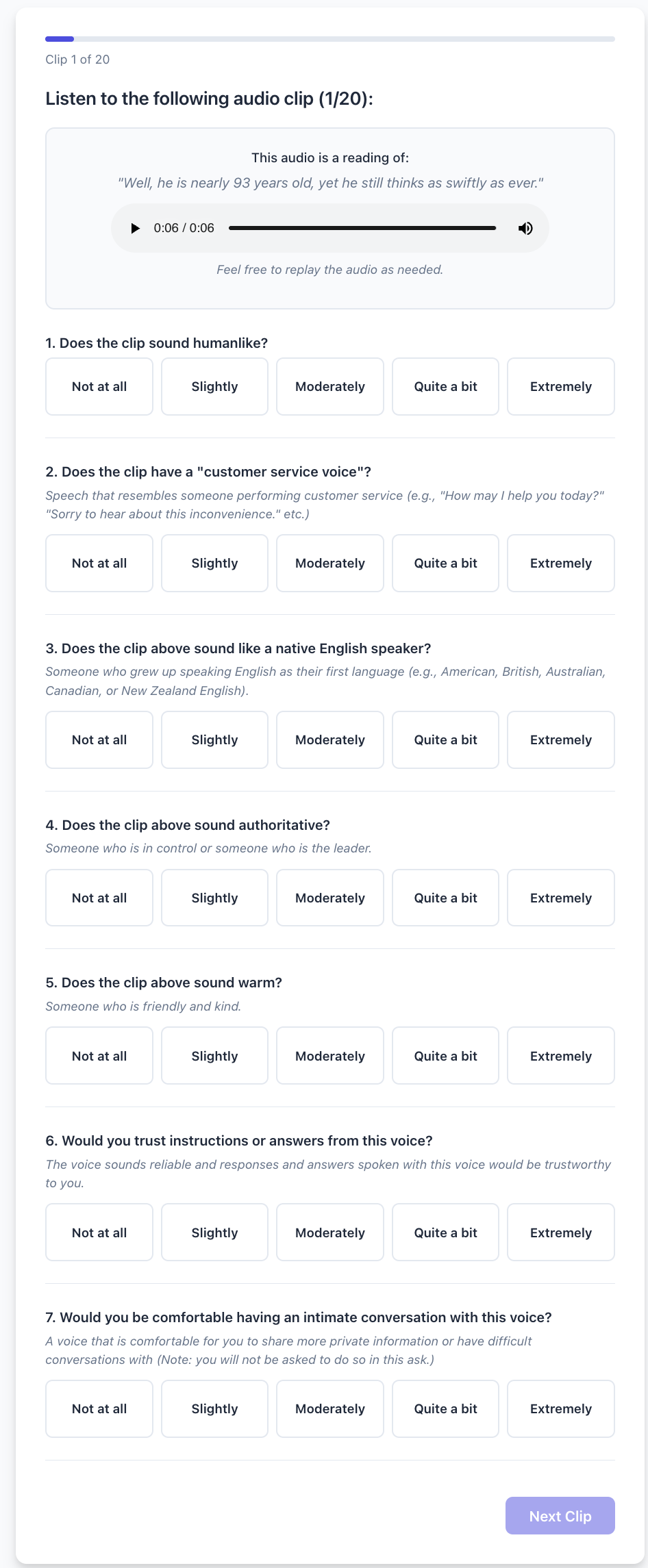}
    \caption{Screenshot of Annotation Task}
    \label{fig:annotation_screenshot}
\end{figure}

\begin{figure}
    \centering
    \includegraphics[width=1\linewidth]{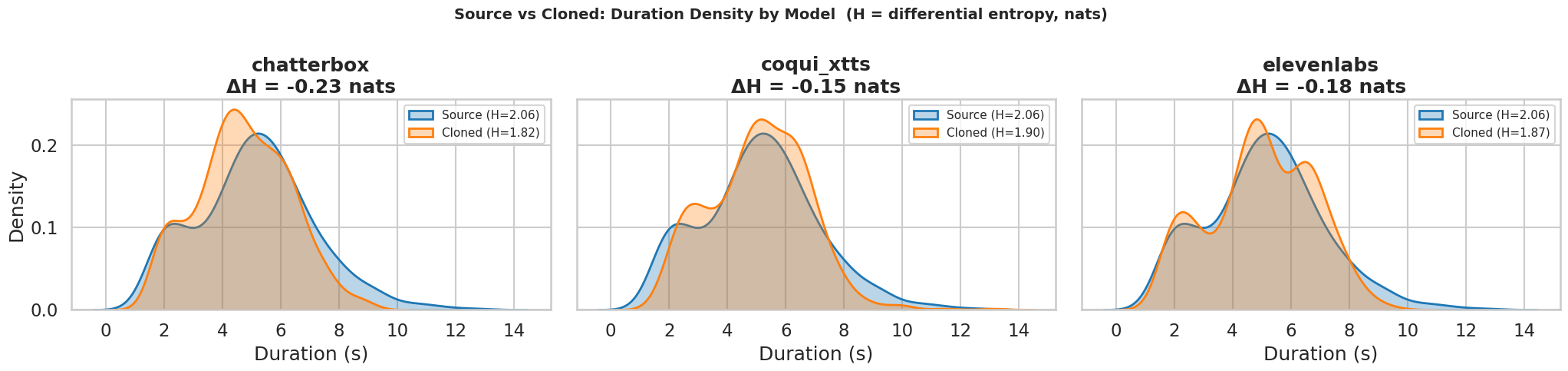}
    \caption{Change in entropy (denoted as nats) for duration distribution between source and cloned recordings}
    \label{fig:duration_entropy}
\end{figure}

\begin{figure}[b]
    \includegraphics[width=\linewidth]{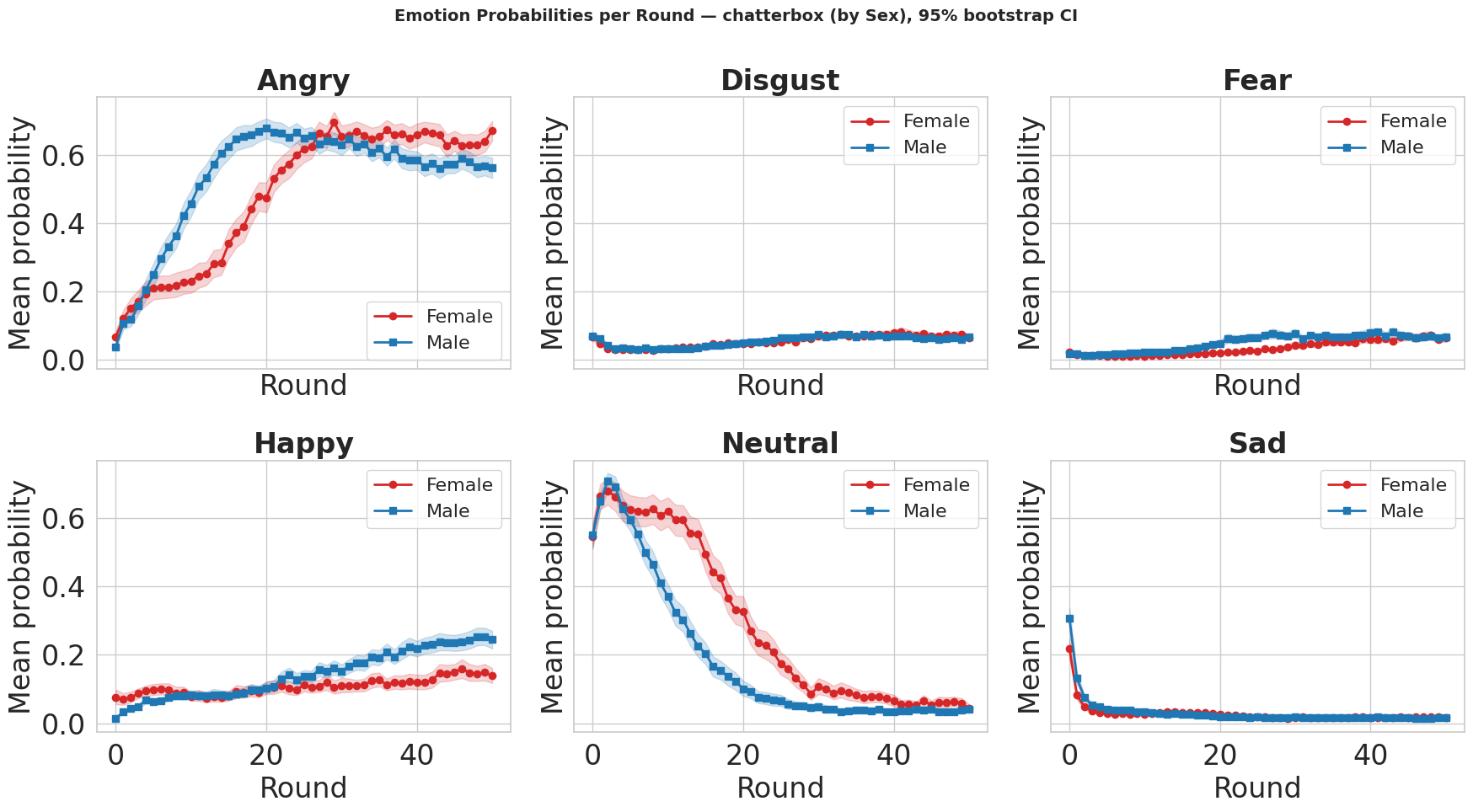}
    \caption{Change predicted emotion across the 50 iterative rounds of cloning, visualized with 95\% confidence interval of the mean calculated via bootstrap resampling.}
    \label{fig:emotion_change}
\end{figure}

\begin{figure}[h!]
\centering
\includegraphics[width=\linewidth]{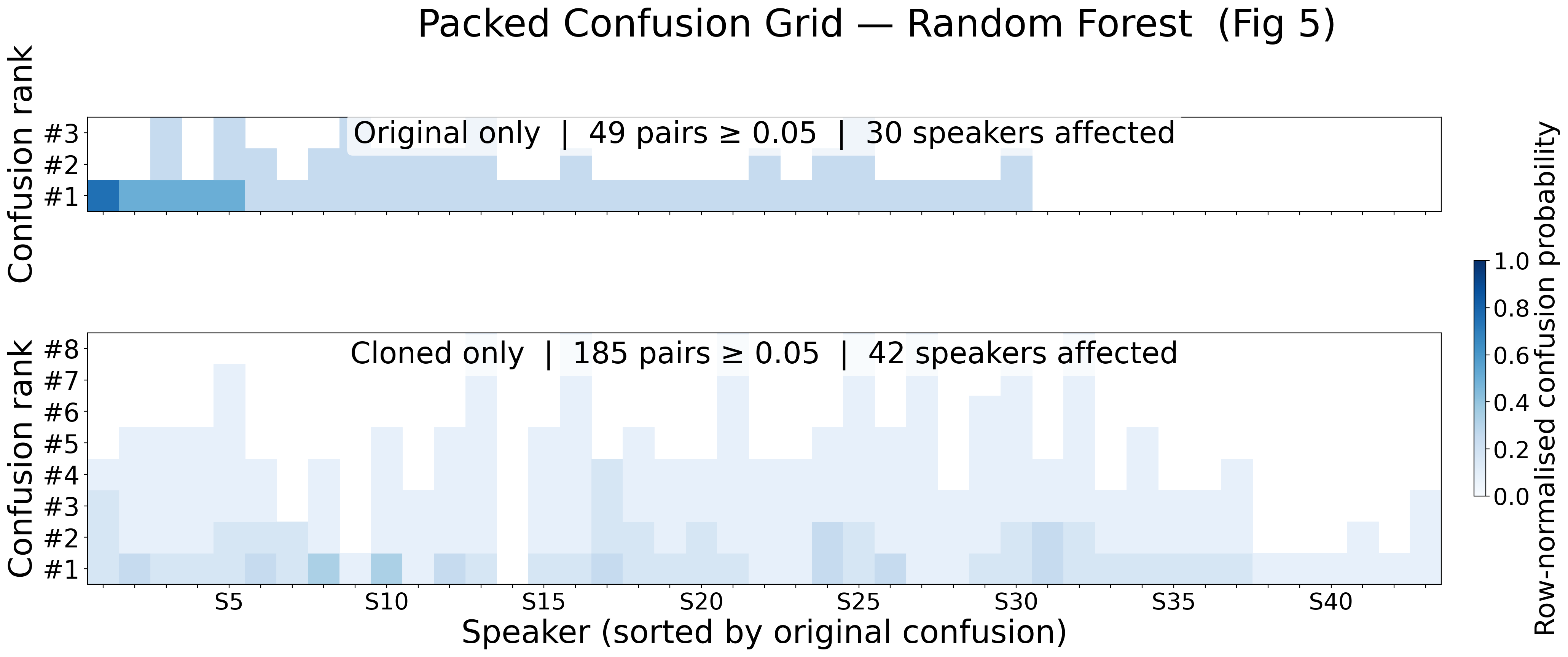}
\caption{Probability distribution on incorrect speakers for source (top) and cloned (bottom) recordings. Each column represents a speaker, and each box in that column represents the probability placed on an incorrect speaker. The more boxes stacked in a column, the more confusion the classifier had in identifying the speaker's true identity. The top figure shows significantly less confusion than the bottom figure, indicating that the classifier confused cloned recordings more often and with more speakers.}
\label{fig:compact_confusion_matrix}
\end{figure}
\clearpage
\subsection{Speaker-Verification Metrics on ECAPA Embeddings}
\label{sec:verification}

\begin{table}[h]
\centering
\caption{Speaker-verification metrics on ECAPA-TDNN embeddings. \emph{Within} is the mean same-speaker cosine similarity and \emph{between} is the mean different-speaker cosine similarity. Cloning closes the gap between mean same-speaker and different-speaker cosine similarity, where a larger gap suggest more separability between speakers. This is most noticeable for \texttt{ElevenLabs}.}

\label{tab:verification}
\begin{tabular}{lccc}
\toprule
\textbf{Condition} & \textbf{Within} & \textbf{Between} & \textbf{Gap} \\
\midrule
Source                       & 0.716 & 0.338 & 0.378 \\
Clone \texttt{ChatterBox}    & 0.610 & 0.303 & 0.307 \\
Clone \texttt{Coqui-XTTS}    & 0.607 & 0.281 & 0.326 \\
Clone \texttt{ElevenLabs V3} & 0.558 & 0.326 & 0.232 \\
\bottomrule
\end{tabular}
\end{table}

\clearpage
\subsection{Accent Controls and Per-Model Effects}
\label{sec:accent_controls}

\paragraph{Speakers who report no accent.} Table \ref{tab:no_accent} limits our experiments to speakers based on their self-reported foreign-accent strength (LEAP-Q). Almost all the effects persist, even among speakers who rated themselves as having no accent. 

\begin{table}[h]
\centering
\caption{Clone$-$source rating differences restricted by self-reported accent strength. $p$-values from permutation tests; $^{*}$ denotes $p<0.05$.}
\label{tab:no_accent}
\begin{tabular}{lccccc}
\toprule
& \multicolumn{2}{c}{\textbf{No accent ($n=16$)}} & \multicolumn{2}{c}{\textbf{No/minimal ($n=30$)}} & \textbf{All ($n=86$)} \\
\cmidrule(lr){2-3}\cmidrule(lr){4-5}\cmidrule(lr){6-6}
\textbf{Trait} & $\Delta$ & $p$ & $\Delta$ & $p$ & $\Delta$ \\
\midrule
Humanlike        & $+0.28^{*}$ & 0.0050 & $+0.39^{*}$ & 0.0002 & $+0.43^{*}$ \\
Customer Service & $+0.23^{*}$ & 0.0104 & $+0.28^{*}$ & 0.0002 & $+0.43^{*}$ \\
Native English   & $+0.24^{*}$ & 0.0138 & $+0.38^{*}$ & 0.0002 & $+0.90^{*}$ \\
Authoritative    & $+0.22^{*}$ & 0.0196 & $+0.30^{*}$ & 0.0002 & $+0.38^{*}$ \\
Warm             & $+0.20^{*}$ & 0.0380 & $+0.28^{*}$ & 0.0002 & $+0.35^{*}$ \\
Trust            & $+0.27^{*}$ & 0.0016 & $+0.36^{*}$ & 0.0002 & $+0.46^{*}$ \\
Intimate Convo   & $+0.13$     & 0.1930 & $+0.22^{*}$ & 0.0022 & $+0.38^{*}$ \\
\bottomrule
\end{tabular}
\end{table}

\paragraph{Mediation by perceived native-English.} Table \ref{tab:mediation} decomposes the total cloning effect into the part transmitted through perceived native-English and the part that is not. The direct effect remains positive for all six traits, with a 95\% CI excluding zero for five of them; warmth is the exception.

\begin{table}[h]
\centering
\caption{Mediation analysis with cloning as treatment and perceived native-English as mediator, for every rated trait except the mediator itself; 1,000 speaker-cluster bootstraps. The direct effect is distinguishable from zero for all traits except warmth, listed last. $c$ is the total effect, $c'$ the direct effect controlling for the mediator.}
\label{tab:mediation}
\begin{tabular}{lccc}
\toprule
\textbf{Trait} & \textbf{Total $c$} & \textbf{Direct $c'$} & \textbf{$c'$ 95\% CI} \\
\midrule
Authoritative    & 0.382 & 0.193 & (0.124, 0.261) \\
Customer Service & 0.429 & 0.215 & (0.132, 0.287) \\
Trust            & 0.459 & 0.134 & (0.068, 0.199) \\
Humanlike        & 0.435 & 0.102 & (0.007, 0.196) \\
Intimate Convo   & 0.377 & 0.081 & (0.013, 0.152) \\
\addlinespace
Warm             & 0.348 & 0.071 & ($-$0.005, 0.150) \\
\bottomrule
\end{tabular}
\end{table}

\begin{table}[h]
\centering
\caption{Clone$-$source effect per trait from crossed random-effects models ($\beta$, 1--5 Likert), with and without per-clip audio-quality covariates, and Cohen's $d$. Random intercepts: speaker, annotator, sentence; TTS model fixed. $n=3{,}000$ ratings; all $p<0.001$.}
\label{tab:mixed_effects}
\small
\begin{tabular}{lccc}
\toprule
\textbf{Trait} & \textbf{$\beta$ (95\% CI)} & \textbf{$\beta$ + quality covariates} & \textbf{Cohen's $d$} \\
\midrule
Warm & $+0.35$ [0.28, 0.41] & $+0.35$ [0.26, 0.43] & 0.30 \\
Authoritative & $+0.38$ [0.32, 0.44] & $+0.37$ [0.29, 0.45] & 0.34 \\
Trust & $+0.46$ [0.40, 0.52] & $+0.44$ [0.36, 0.52] & 0.41 \\
Humanlike & $+0.43$ [0.36, 0.51] & $+0.40$ [0.30, 0.50] & 0.35 \\
Customer Service & $+0.43$ [0.36, 0.50] & $+0.44$ [0.35, 0.53] & 0.36 \\
Native English & $+0.90$ [0.82, 0.98] & $+0.78$ [0.68, 0.88] & 0.66 \\
Intimate Convo & $+0.38$ [0.32, 0.44] & $+0.36$ [0.28, 0.44] & 0.32 \\
\bottomrule
\end{tabular}
\end{table}

\paragraph{Per-model effects.} Table \ref{tab:mixed_effects_bymodel} reports the crossed random-effects estimates separately for each system. The direction is consistent across all three, with \texttt{ElevenLabs V3} generally showing the largest shifts.

\begin{table}[h]
\centering
\caption{Clone$-$source effect ($\beta$, Likert 1--5) by TTS system, from crossed random-effects models with random intercepts for speaker, annotator, and sentence. $^{*}p<0.05$, $^{**}p<0.01$, $^{***}p<0.001$.}
\label{tab:mixed_effects_bymodel}
\begin{tabular}{lccc}
\toprule
\textbf{Trait} & \textbf{\texttt{ChatterBox}} & \textbf{\texttt{Coqui-XTTS}} & \textbf{\texttt{ElevenLabs V3}} \\
\midrule
Warm             & $+0.29$ [0.17, 0.40]$^{***}$ & $+0.23$ [0.12, 0.34]$^{***}$ & $+0.50$ [0.40, 0.61]$^{***}$ \\
Authoritative    & $+0.40$ [0.29, 0.52]$^{***}$ & $+0.27$ [0.16, 0.38]$^{***}$ & $+0.47$ [0.37, 0.57]$^{***}$ \\
Trust            & $+0.44$ [0.33, 0.55]$^{***}$ & $+0.29$ [0.19, 0.40]$^{***}$ & $+0.63$ [0.53, 0.72]$^{***}$ \\
Humanlike        & $+0.37$ [0.22, 0.52]$^{***}$ & $+0.35$ [0.23, 0.48]$^{***}$ & $+0.56$ [0.44, 0.67]$^{***}$ \\
Customer Service & $+0.44$ [0.32, 0.57]$^{***}$ & $+0.25$ [0.13, 0.36]$^{***}$ & $+0.59$ [0.48, 0.70]$^{***}$ \\
Native English   & $+0.69$ [0.54, 0.84]$^{***}$ & $+0.94$ [0.80, 1.07]$^{***}$ & $+1.01$ [0.88, 1.13]$^{***}$ \\
Intimate Convo   & $+0.27$ [0.17, 0.38]$^{***}$ & $+0.29$ [0.18, 0.40]$^{***}$ & $+0.53$ [0.43, 0.64]$^{***}$ \\
\bottomrule
\end{tabular}
\end{table}

\end{document}